\documentclass[letterpaper,twocolumn,10pt]{article}
\usepackage{usenix}

\usepackage{tikz}
\usepackage{amsmath}
\usepackage{xcolor}

\usepackage{filecontents}

\usepackage{array}
\usepackage{booktabs}

\usepackage{enumitem}

\usepackage{amssymb}
\usepackage{pifont}

\usepackage{wasysym}
\usepackage{xspace}

\usepackage{multirow}
\usepackage{colortbl}
\usepackage[most]{tcolorbox}
\usepackage{makecell}
\newcommand{\peiran}[1]{{\textcolor{red}{[\textit{Peiran: #1}]}}}

\usepackage{subcaption}

\newcolumntype{M}[1]{>{\centering\arraybackslash}p{#1}}
\newcommand{\sssec}[1]{\vspace*{0.05in}\noindent\textbf{#1}}

\usepackage{fontawesome}
\usepackage{hyperref}
\usepackage{url}

\makeatletter
\newcommand{\github}[1]{%
   \href{#1}{\faGithub}%
}
\makeatother

\usepackage[normalem]{ulem}
\usepackage[table,xcdraw, svgnames]{xcolor}
\usepackage[normalem]{ulem}
\useunder{\uline}{\ul}{}

\DeclareRobustCommand*\circled[1]{\tikz[baseline=(char.base)]{ \node[shape=circle,draw,color=white,fill=black,inner sep=0.5pt] (char){#1};}}
\usepackage{tikz}
\DeclareRobustCommand*\circledRed[1]{\tikz[baseline=(char.base)]{ \node[shape=circle,draw,color=white,fill=red!40!black,inner sep=0.5pt] (char){#1};}}
\DeclareRobustCommand*\circledGreen[1]{\tikz[baseline=(char.base)]{ \node[shape=circle,draw,color=white,fill=green!30!black,inner sep=0.5pt] (char){#1};}}
\DeclareRobustCommand*\circledOrange[1]{\tikz[baseline=(char.base)]{ \node[shape=circle,draw,color=white,fill=orange!50!black,inner sep=0.5pt] (char){#1};}}
\DeclareRobustCommand*\circledBlue[1]{\tikz[baseline=(char.base)]{ \node[shape=circle,draw,color=white,fill=blue!40!black,inner sep=0.5pt] (char){#1};}}

\newcommand{\No}{\raisebox{-.25em}{\includegraphics[width=1em]{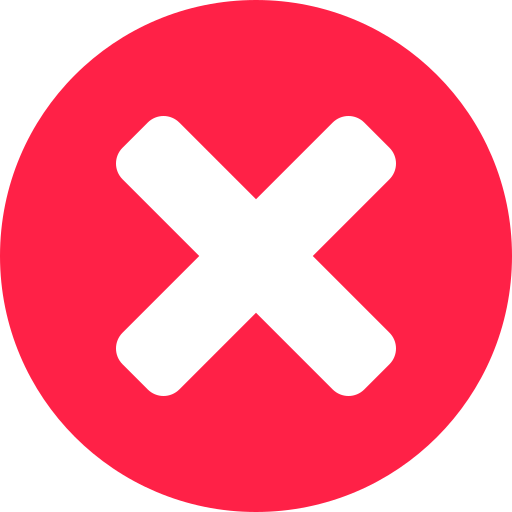}}}
\newcommand{\BlackBox}{\raisebox{-.25em}{\includegraphics[width=1em]{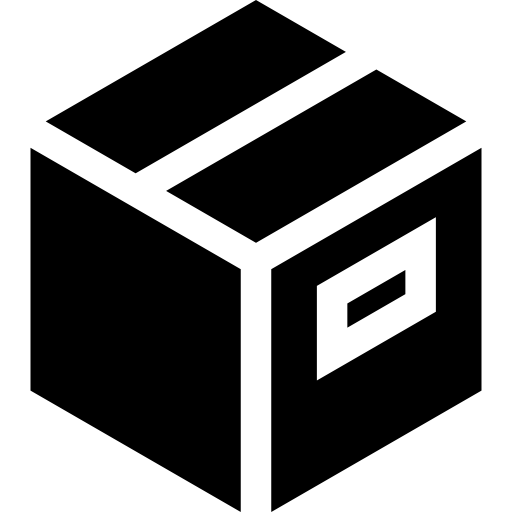}}}
\newcommand{\GreyBox}{\raisebox{-.25em}{\includegraphics[width=1em]{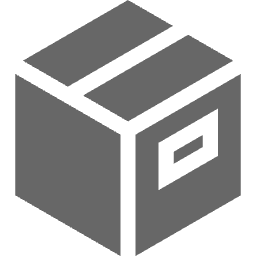}}}
\newcommand{\WhiteBox}{\raisebox{-.25em}{\includegraphics[width=1em]{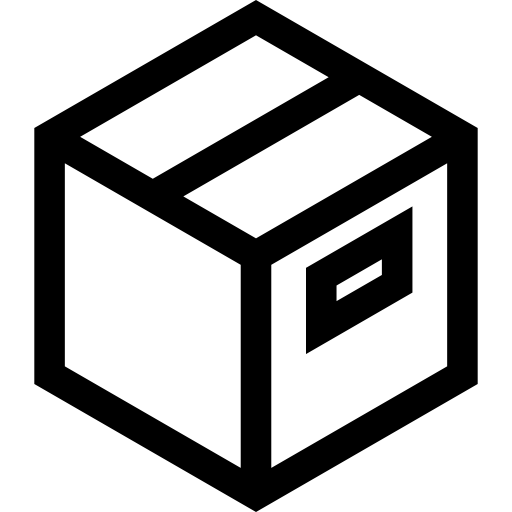}}}

\newcommand{\Visible}{\textcolor{DarkBlue}{\faEye}} 
\newcommand{\InVisible}{\textcolor{DarkRed}{\faEyeSlash}}

\newcommand{\bench}{\textsc{AgentPI}\xspace}

\usepackage{algorithm}
\usepackage{algorithmic}

\usepackage{verbatim}

\newcommand{\TextInput}{\raisebox{-.25em}{\includegraphics[width=1em]{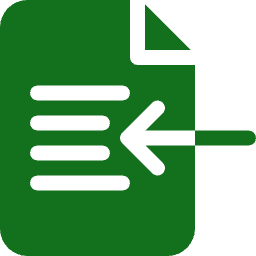}}}
\newcommand{\TextOutput}{\raisebox{-.25em}{\includegraphics[width=1em]{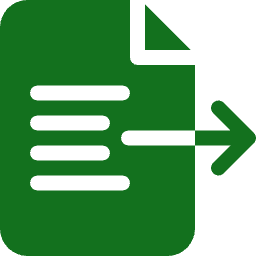}}}
\newcommand{\ModelIR}{\raisebox{-.25em}{\includegraphics[width=1em]{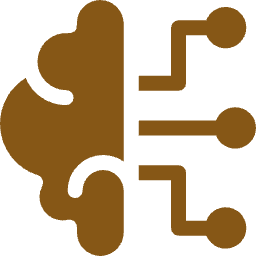}}}
\newcommand{\ModelParam}{\raisebox{-.25em}{\includegraphics[width=1em]{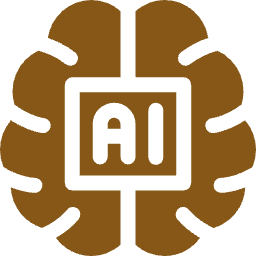}}}
\newcommand{\ToolCall}{\raisebox{-.25em}{\includegraphics[width=1em]{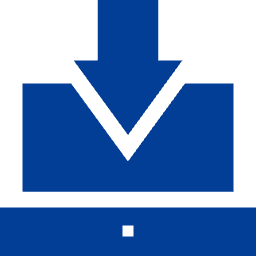}}}
\newcommand{\ToolObs}{\raisebox{-.25em}{\includegraphics[width=1em]{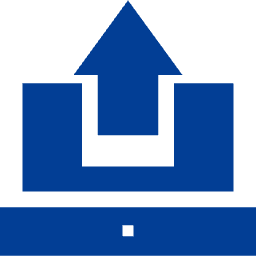}}}
\newcommand{\SystemPrompt}{\raisebox{-.25em}{\includegraphics[width=1em]{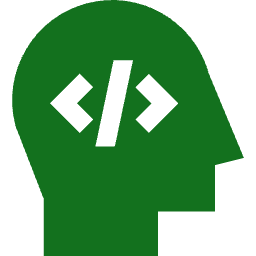}}}
\newcommand{\UserPrompt}{\raisebox{-.25em}{\includegraphics[width=1em]{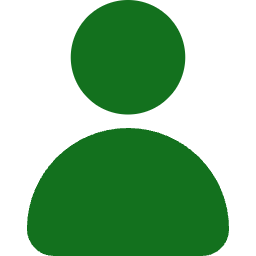}}}
\newcommand{\Environment}{\raisebox{-.25em}{\includegraphics[width=1em]{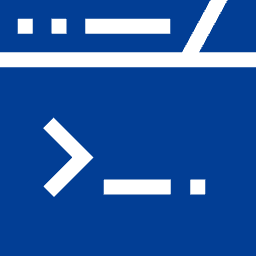}}}
\newcommand{\AgentArch}{\raisebox{-.25em}{\includegraphics[width=1em]{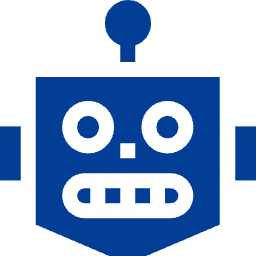}}}
\newcommand{\ContextMemory}{\raisebox{-.25em}{\includegraphics[width=1em]{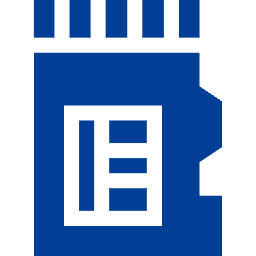}}}

\DeclareRobustCommand*\circledLightRedProperty[1]{\tikz[baseline=(char.base)]{ \node[draw=black,shape=circle,draw,color=black,fill=red!50!white,inner sep=0pt, minimum size=1em] (char){#1};}}
\DeclareRobustCommand*\circledLightOrangeProperty[1]{\tikz[baseline=(char.base)]{ \node[draw=black,shape=circle,draw,color=black,fill=orange!50!white,inner sep=0pt, minimum size=1em] (char){#1};}}
\DeclareRobustCommand*\circledLightGreenProperty[1]{\tikz[baseline=(char.base)]{ \node[draw=black,shape=circle,draw,color=black,fill=green!50!white,inner sep=0pt, minimum size=1em] (char){#1};}}
\DeclareRobustCommand*\circledLightBlueProperty[1]{\tikz[baseline=(char.base)]{ \node[draw=black,shape=circle,draw,color=black,fill=blue!50!white,inner sep=0pt, minimum size=1em] (char){#1};}}

\newcommand{\ConfProperty}{\circledLightRedProperty{C}}
\newcommand{\InteProperty}{\circledLightOrangeProperty{I}}
\newcommand{\AvaiProperty}{\circledLightGreenProperty{A}}
\newcommand{\GeneralProperty}{\circledLightBlueProperty{G}}

\newcommand{\HighCompa}{\raisebox{-.25em}{\includegraphics[width=1em]{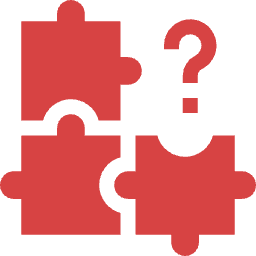}}}
\newcommand{\MedCompa}{\raisebox{-.25em}{\includegraphics[width=1em]{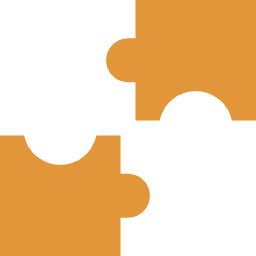}}}
\newcommand{\LowCompa}{\raisebox{-.25em}{\includegraphics[width=1em]{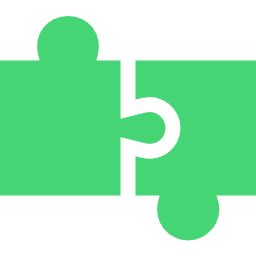}}}

\newcommand{\ExplainIR}{\raisebox{-.25em}{\includegraphics[width=1em]{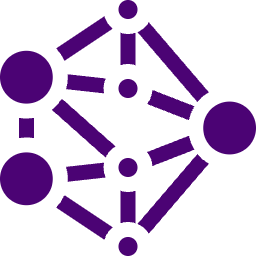}}}
\newcommand{\ExplainLLM}{\raisebox{-.25em}{\includegraphics[width=1em]{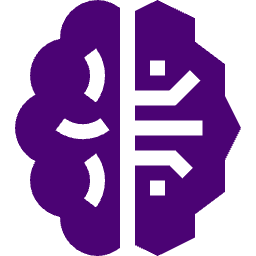}}}
\newcommand{\ExplainHuman}{\raisebox{-.25em}{\includegraphics[width=1em]{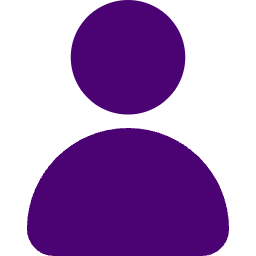}}}
\newcommand{\ExplainSem}{\raisebox{-.25em}{\includegraphics[width=1em]{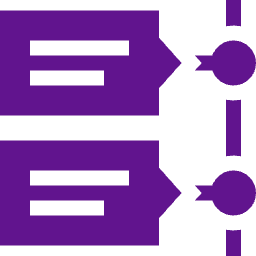}}}

\newcommand{\FullAuto}{\raisebox{-.25em}{\includegraphics[width=1em]{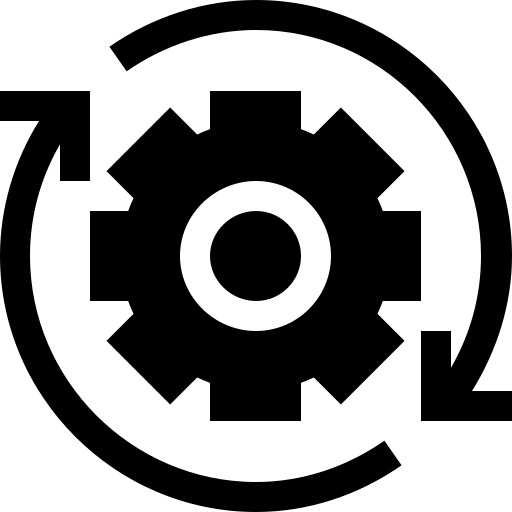}}}
\newcommand{\SemiAuto}{\raisebox{-.25em}{\includegraphics[width=1em]{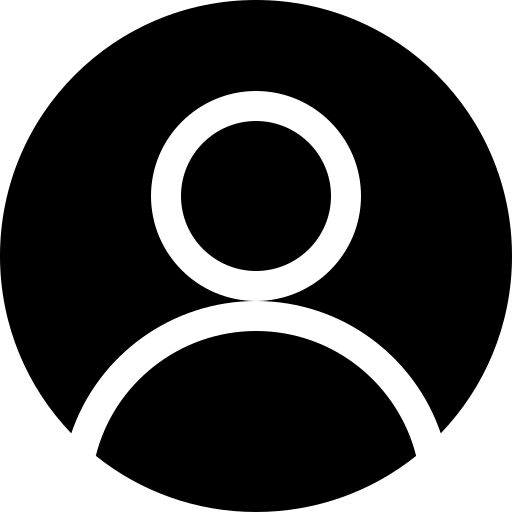}}}

\usepackage{diagbox}

\newcommand{\UtilityStatic}{\circledLightRedProperty{S}}
\newcommand{\UtilityDynamic}{\circledLightGreenProperty{D}}
\newcommand{\UtilityHuman}{\circledLightOrangeProperty{H}}

\newcommand{\AtkDirect}{\raisebox{-.25em}{\includegraphics[width=1em]{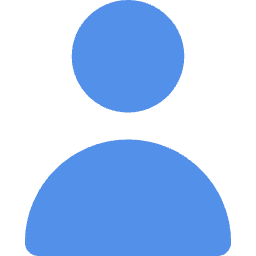}}}
\newcommand{\AtkIndirect}{\raisebox{-.25em}{\includegraphics[width=1em]{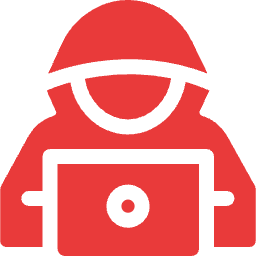}}}
\newcommand{\AtkSupply}{\raisebox{-.25em}{\includegraphics[width=1em]{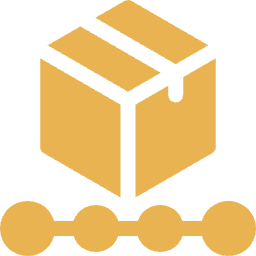}}}

\newcommand{\VictimLLM}{\raisebox{-.25em}{\includegraphics[width=1em]{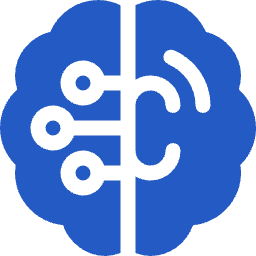}}}
\newcommand{\VictimApp}{\raisebox{-.25em}{\includegraphics[width=1em]{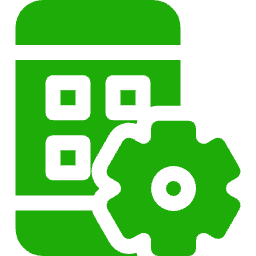}}}
\newcommand{\VictimAgent}{\raisebox{-.25em}{\includegraphics[width=1em]{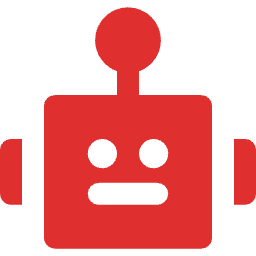}}}

\newcommand{\InVisibleSemantics}{Semantics}
\newcommand{\InVisibleContext}{Context}
\newcommand{\InVisibleEncoding}{Encoding}
\newcommand{\InVisibleVision}{Vision}

\newcounter{definition}

{%
    \refstepcounter{definition}%
    \textbf{\textit{Definition} \thedefinition. }\space %
}%
{%
}
\newcounter{problem}

{%
    \refstepcounter{problem}%
    \textbf{Problem Definition.}\space%
}%
{%
}
\newcounter{finding}
\newcounter{openrq}
\newcounter{takeaway}
\newcommand{\takeaways}[1]{
\vspace{1pt}
\noindent
\begin{tcolorbox}[ enhanced, 
    breakable,
    boxrule=1pt, 
    arc=4pt,
    left=2pt,
    right=2pt,
    bottom=2pt,
    top=2pt,
    colback=gray!4,        
    colframe=gray!1!black,
    drop shadow=black!50!white, 
    rounded corners]
\noindent 
\refstepcounter{takeaway}
\textbf{Takeaway \Roman{takeaway}.} \small
{#1}
\end{tcolorbox}
}

\newcounter{rootcause}
\newcounter{oproblem}
\newcommand{\openquestion}[1]{
\vspace{1pt}
\noindent
\begin{tcolorbox}[ 
    enhanced, 
    breakable,
    boxrule=1pt, 
    arc=4pt,
    left=2pt,
    right=2pt,
    bottom=2pt,
    top=2pt,
    colback=gray!4,        
    colframe=gray!1!black,
    drop shadow=black!50!white, 
    rounded corners,
    boxrule=0pt,                
    frame hidden,               
    borderline={1pt}{0pt}{black, dashed},
]
\noindent 
\refstepcounter{oproblem}
\textbf{Open Problem \Roman{oproblem}.} \small
{#1}
\end{tcolorbox}
}

\newcommand{\ignore}[1]{}

\newcommand{\xf}[1]{{{\textcolor{blue}{[Xinfeng: #1]}}}}

\usepackage{threeparttable}

\usepackage{listings}
\usepackage{xcolor}
\colorlet{punct}{red!60!black}
\definecolor{delim}{RGB}{20,105,176}
\colorlet{numb}{magenta!60!black}

\lstdefinelanguage{json}{
    basicstyle=\normalfont\ttfamily\footnotesize,
    numbers=left,
    numberstyle=\scriptsize,
    stepnumber=1,
    numbersep=8pt,
    showstringspaces=false,
    breaklines=true,
    frame=lines,
    literate=
     *{0}{{{\color{numb}0}}}{1}
      {1}{{{\color{numb}1}}}{1}
      {2}{{{\color{numb}2}}}{1}
      {3}{{{\color{numb}3}}}{1}
      {4}{{{\color{numb}4}}}{1}
      {5}{{{\color{numb}5}}}{1}
      {6}{{{\color{numb}6}}}{1}
      {7}{{{\color{numb}7}}}{1}
      {8}{{{\color{numb}8}}}{1}
      {9}{{{\color{numb}9}}}{1}
      {:}{{{\color{punct}{:}}}}{1}
      {,}{{{\color{punct}{,}}}}{1}
      {\{}{{{\color{delim}{\{}}}}{1}
      {\}}{{{\color{delim}{\}}}}}{1}
      {[}{{{\color{delim}{[}}}}{1}
      {]}{{{\color{delim}{]}}}}{1},
}

\begin{document}

\date{}

\title{The Landscape of Prompt Injection Threats in LLM Agents: From Taxonomy to Analysis}


\author{
{\rm Peiran Wang}\\ UCLA \and
{\rm Xinfeng Li}\\ NTU \and
{\rm Chong Xiang}\\ NVIDIA \and
{\rm Jinghuai Zhang}\\ UCLA \and
{\rm Ying Li}\\ UCLA \and
{\rm Lixia Zhang}\\ UCLA \and
{\rm Xiaofeng Wang}\\ NTU \and
{\rm Yuan Tian}\\ UCLA
}

\maketitle

\begin{abstract}

The evolution of Large Language Models (LLMs) has resulted in a paradigm shift towards autonomous agents, necessitating robust security against Prompt Injection (PI) vulnerabilities where untrusted inputs hijack agent behaviors. 
This SoK presents a comprehensive overview of the PI landscape, covering attacks, defenses, and their evaluation practices.
Through a systematic literature review and quantitative analysis, we establish taxonomies that categorize PI attacks by payload generation strategies (heuristic vs. optimization) and defenses by intervention stages (text, model, and execution levels).
Our analysis reveals a key limitation shared by many existing defenses and benchmarks: they largely overlook \emph{context-dependent tasks}, in which agents are authorized to rely on runtime environmental observations to determine actions.
To address this gap, we introduce \bench, a new benchmark designed to systematically evaluate agent behavior under context-dependent interaction settings.
Using \bench, we empirically evaluate representative defenses and show that no single approach can simultaneously achieve high trustworthiness, high utility, and low latency.
Moreover, we show that many defenses appear effective under existing benchmarks by suppressing contextual inputs, yet fail to generalize to realistic agent settings where context-dependent reasoning is essential.
This SoK distills key takeaways and open research problems, offering structured guidance for future research and practical deployment of secure LLM agents.

\end{abstract}

\section{Introduction}\label{sec:introduction}

Recent advances have enabled Large Language Model (LLM) agents to interact with external tools and environments, substantially expanding their capabilities~\cite{yang2023auto, hong2023metagpt, wang2024mobile}.
However, the LLM agents are vulnerable to \textit{Prompt Injection} (PI) attacks, in which untrusted inputs manipulate the backbone LLM's output to hijack the agent behavior~\cite{liu2024automatic, pasquini2024hacking}. 
Such attacks can lead to severe consequences, including unauthorized remote computer use \cite{vandevanter2025prompt}, sensitive data leakage \cite{hackerone2024prompt}, etc.


Thus, many works (78 papers we collected until Oct. 20, 2025) have been proposed to study such threats. 
On the attack side, for example, optimization-based techniques utilize fuzzing \cite{yu2024promptfuzz} or gradient guidance \cite{liu2024automatic, pandya2025may} to generate stealthy payloads, yet they often face challenges regarding query efficiency or require white-box access. 
On the defense side, some detection approaches employing external LLMs \cite{shi2025promptarmor, liu2025secinfer} show promise in identifying payload segments, but they introduce additional computational overhead. 
Similarly, isolation works \cite{wu2025isolategpt, wu2024system} aim to contain threats by separating control and data flow, although they may severely compromise the agent's utility in complex tasks.

Given the rapid growth of this field, a SoK is crucial to understand the landscape.
This SoK aims to:
(1) bridge knowledge gaps by analyzing PI attack and defense approaches; 
(2) offer critical insights into the core paradigms, strengths, and limitations of current work;
(3) propose promising open problems to motivate future research directions.
In this paper, we first establish a taxonomy of existing PI attacks grounded in attack payload generation methods.
The taxonomy provides a trend analysis of payload generation, attack threat models, attacker capability, and payload visibility. 
We found that attacks have gradually evolved from impractical attacks under white-box, base LLM victim settings to more practical settings under black-box, LLM agent victim settings.

Next, we provide a taxonomy of existing PI defenses categorized by defense intervention stages.
This taxonomy provides a comparative analysis of intervention stages, defense capability, explainability, and costs, revealing a multifaceted landscape.
Furthermore, we identify several takeaways and open problems to motivate future research from this taxonomy.
We identified that there is no definitive ``perfect'' defense to meet the high trustworthiness (security+explainability), high utility, and low latency simultaneously.
For instance, the LLM-involved methods lack reliability in terms of explainability \cite{wang2025agentarmor, shi2025promptarmor}, while human-involved methods bring additional latency costs \cite{wu2025isolategpt, shi2025progent}.
We propose several important research directions to advance PI, including the integration of availability defenses, fine-grained access control for attention probe for explainability, etc.
We also identify a key limitation shared by existing defenses: they lack consideration of \emph{context-dependent tasks}.

To address the lack of consideration on \emph{context-dependent tasks}, we propose \bench benchmark. 
Real-world agents frequently rely on runtime environmental observations, such as following configuration files~\cite{xie2024osworld, jimenez2023swe} or executing conditional logic (e.g. ``If-Else'' structure)~\cite{wang2025agentarmor}. 
This reliance creates \emph{context-dependent tasks} and an attack surface for \textit{context-aware attacks}, in which adversaries manipulate environmental inputs to corrupt reasoning. 
However, existing benchmarks \cite{debenedetti2024agentdojo, zhang2024asb, zhan2024injecagent} predominantly focus on tasks fully specified by prompts. 
This focus inadvertently favors defenses that isolate or reject contextual inputs, leading to an overestimation of their utility in realistic scenarios where context is essential for planning. 
To bridge this gap, \bench systematizes 5 context-dependent tasks and their corresponding attacks.

Finally, to validate our taxonomy and theoretical analysis, we conducted comprehensive experiments on 8 defenses, ranging from text-level filters to execution-level monitors. 
We evaluated these mechanisms using a multi-dimensional metric system that quantifies the trade-offs between security effectiveness (attack success rate), agent utility, and computational cost (time and tokens). 
Our empirical analysis reveals that there is currently no tested defense that simultaneously achieves high trustworthiness, high utility, and low latency. 
Specifically, we find that while execution-level defenses can enforce strict action boundaries, they often incur prohibitive computational overhead, up to 3x the baseline cost, or resort to aggressive ``early refusal'' strategies that render the agent unusable for complex tasks. 
Conversely, text-level defenses, while efficient, fail to provide robust security guarantees against sophisticated payload manipulations.

In summary, we have the following key contributions:
\begin{itemize}[noitemsep, topsep=1pt, leftmargin=*]
    \item \textbf{Systematic taxonomy of attacks and defenses}: We provide a comprehensive taxonomy and comparative analysis of 78 papers, encompassing both attacks and defenses. We categorize attacks based on payload generation methodologies (heuristic vs. optimization) and defenses by intervention stages (text, model, and execution levels). This systematization offers a unified view of the evolving PI landscape, from manual templates to automated injections, and the corresponding mitigation strategies.
    \item \textbf{The \bench benchmark and empirical evaluation}: To address the critical lack of evaluation on context-dependent tasks, we introduce \bench benchmark, designed to assess defenses' performance in such tasks. We evaluate 8 defenses and identify a fundamental trilemma among trustworthiness, utility, and latency.
    \item \textbf{Insights and future roadmap}: We explore the latest research trends, identify key challenges, and propose future directions based on 9 takeaways and 4 open problems. Through quantitative and qualitative analysis, we highlight gaps in current works where there is no ``perfect'' defense to meet high trustworthiness, high utility, and low latency, and extract open problems to guide future research in PI.
\end{itemize}

\begin{figure*}[!htbp]
    \centering
    \includegraphics[width=\linewidth]{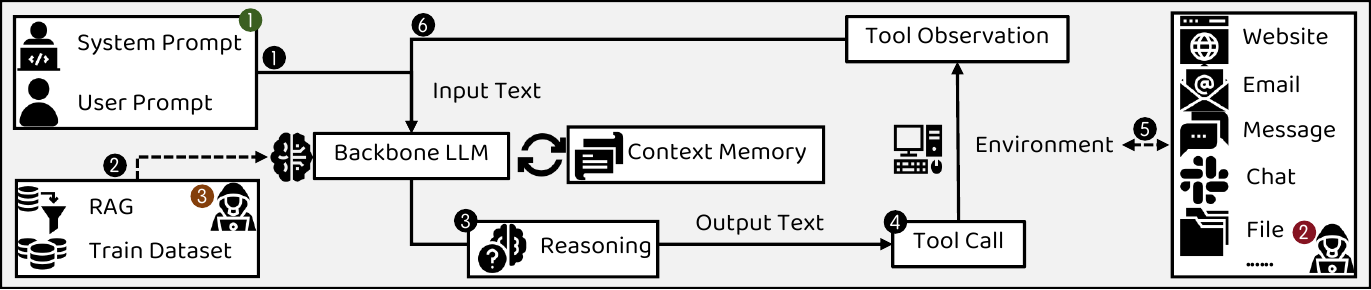}
    \caption{Overview of the LLM agent execution loops and inputs to the agent.}
    \label{fig:agent_definition}
\end{figure*}

\section{Preliminary and Problem Setup}\label{sec:preliminary}



In this section, we first define the system model of LLM agents in \S\ref{sec:preliminary:lifecycle}.
Then, we present the definition of prompt injection in LLM agents in \S\ref{sec:preliminary:PI}.

\subsection{LLM Agents}\label{sec:preliminary:lifecycle}

To provide a system model for this paper, we first introduce the typical execution loop of the LLM agentic system, with 6 iterative steps as illustrated in Fig. \ref{fig:agent_definition}: 
\begin{enumerate}[noitemsep, topsep=1pt, leftmargin=*, label={\circled{\arabic*}}]
    \item \textbf{Receive prompts}. Firstly, LLM agents receive system prompts defined by the agentic developer and the user prompts written by the users.
    \item \textbf{Retrieve RAG (optional)}. Some of the agents are integrated with the Retrieval-Augmented Generation (RAG) database to fetch relevant, up-to-date data from external knowledge sources before generating the response.
    \item \textbf{Reasoning (optional)}. Next, the agents go through a chain-of-thought process to reason about the plans or the next step to execute. This step is optional, since some agent developers tend to use function calling \cite{openai-function-call}.
    \item \textbf{Generate tool call}. Then, based on previous context memory, the backbone LLM of the agents generates the next step's tool call (the tool name and tool parameters).
    \item \textbf{Tool execution}. The tool call is forwarded to the executive environments bound with the LLM agents to execute.
    \item \textbf{Tool observation return (loop back to \circled{2})}. At last, the environment returns tool execution results (called tool observation) to the LLM agents, integrating into the context memory, and looping back to \circled{2}.
\end{enumerate}

\subsection{Prompt Injection in LLM Agents}\label{sec:preliminary:PI}


\sssec{Input to the LLM agents.}
Within this loop, the LLM agents take inputs from various sources, along with their influence on the actions of LLM agents.
Unlike computer programs, which provide isolated input interfaces for diverse inputs, LLM agents manage all the inputs in a single context memory.
Different inputs are segmented with separators (e.g., ``[SYSTEM]'', ``[DATA]'', etc.) in the context memory.
We categorize the input into 3 types as shown in Fig. \ref{fig:agent_definition}:
\begin{enumerate}[noitemsep, topsep=1pt, leftmargin=*, label={}]
\item \circledGreen{1} \textbf{Trusted prompt input}. The system prompt and user prompt constitute the basic input of an LLM agent. Among them, the system prompt is defined by the agentic developer, while the user prompt is issued by the user. The prompts are generally considered trusted in LLM agents.
\item \circledRed{2} \textbf{Untrusted tool observation}. Another type of input is tool observation, originating from the tool execution process within the environment. Since the executive environment is not fully controlled by the agent, the tool observation is considered to be the untrusted input.
\item \circledOrange{3} \textbf{Supply-chain dataset input}. In addition to the 2 common-seen input types, the supply-chain data, including the retrieval content of RAG and the training dataset, are also considered. Most works treat these inputs as trusted; however, some threat models account for attack surfaces from the 2 data sources \cite{zhang2024hijackrag, shao2024enhancing, zhang2024towards}. Thus, we label these 2 inputs as partially trusted.
\end{enumerate}
These inputs have different trust levels and co-exist in a single context memory of the LLM agent to affect its behaviors.

\sssec{Principles of prompt injection}. 
The prompt injection vulnerability arises from the lack of privilege isolation between different trust levels' inputs. 
Despite textual delimiters, the model processes the context as a unified semantic sequence, creating ambiguity where untrusted inputs can mimic trusted input to manipulate the agent's behaviors. 
This results in \textit{attention competition}: adversarial payloads manipulate the self-attention mechanism, shifting attention weights away from system prompts toward malicious inputs \cite{hung2025attention, zhong2025attention}. 
Furthermore, this leads to unauthorized privilege escalation, enabling the hijacking of the agent's control flow.

\section{Taxonomy of Attacks}\label{sec:attack_survey}

\sssec{Selection methodology}.
We conducted a systematic literature search focused on prompt injection attacks using two search queries: ``prompt injection attacks'' and ``LLM agent attacks''.
The search was performed on \href{https://scholar.google.com/}{Google Scholar} manually, and excluding result papers about jailbreaking LLM agents, general LLM agent safety, etc.
The selection process ended on Oct. 20, 2025, with \textit{37 prompt injection attack papers} selected.
In addition, during our selection, the attack papers on general LLM and specific LLM applications (LLM translation, etc.) instead of agents are selected as well, since the attacks can be integrated into LLM agents as well.

\sssec{Taxonomy methodology}.
We systematically categorized the collected prompt injection attack papers based on the payload generation methods.
We identified two types: heuristic-based and optimization-based in \S\ref{sec:attack:method}.

\sssec{Analysis methodology}.
After systematic taxonomy analysis, we discuss the core attack paradigm across all attacks, including the threat models (attack surfaces, victims, and goals), the attacker capabilities, and the visibility of payloads in \S\ref{sec:attack:dimension}.

\begin{table*}[!t]
\centering
\begin{threeparttable}
\footnotesize
\caption{Systematization of prompt injection attacks, categorized into optimization and heuristic as illustrated in \S\ref{sec:attack:method}.}
\label{tab:attack-taxonomy}
\begin{tabular}{|c|c|c|l|l|l|l|c|}
\hline\hline
Category & Method\S\ref{sec:attack:method} & Surface\S\ref{sec:attack:dimension} & \multicolumn{1}{c|}{Victim\S\ref{sec:attack:dimension}} & \multicolumn{1}{c|}{Goal\S\ref{sec:attack:dimension}} & \multicolumn{1}{c|}{Capability\S\ref{sec:attack:dimension}} & Visibility\S\ref{sec:attack:dimension} & Ref. \\ \hline\hline
\multirow{14}{*}{Optimization} & \multirow{6}{*}{Gradient} & \AtkDirect & \VictimLLM:Base & \InteProperty:Goal Hijack & \WhiteBox:Gradients & \InVisible:\InVisibleSemantics & \cite{liu2024automatic} \\ \cline{3-8} 
 &  & \AtkDirect & \VictimLLM:Base & \GeneralProperty:Fingerprint & \WhiteBox:Logits & \InVisible:\InVisibleSemantics & \cite{hu2025fingerprinting} \\ \cline{3-8} 
 &  & \AtkDirect & \VictimLLM:Base & \InteProperty:Goal Hijack & \WhiteBox:Gradients & \InVisible:\InVisibleSemantics & \cite{pasquini2024neural} \\ \cline{3-8} 
 &  & \AtkDirect & \VictimApp:Evaluator & \InteProperty:Score Tamper & \WhiteBox:Logits & \InVisible:\InVisibleSemantics & \cite{shi2024optimization} \\ \cline{3-8} 
 &  & \AtkIndirect & \VictimLLM:Finetuned & \GeneralProperty:Defend Bypass & \WhiteBox:Attention & \InVisible:\InVisibleSemantics & \cite{pandya2025may} \\ \cline{3-8} 
 &  & \AtkIndirect & \VictimLLM:RAG & \InteProperty:Goal Hijack & \WhiteBox:Gradients & \InVisible:\InVisibleContext & \cite{zhang2024hijackrag} \\ \cline{2-8} 
 & \multirow{6}{*}{Genetic} & \AtkDirect & \VictimLLM:Base & \InteProperty:Goal Hijack & \BlackBox:Query-free & \InVisible:\InVisibleSemantics & \cite{zhang2024goal} \\ \cline{3-8} 
 &  & \AtkDirect & \VictimLLM:Base & \InteProperty:Goal Hijack & \BlackBox:Success Boolean & \Visible & \cite{yu2024promptfuzz} \\ \cline{3-8} 
 &  & \AtkIndirect & \VictimAgent:Tabular & \InteProperty:Goal Hijack & \BlackBox:Shadow Agent & \InVisible:\InVisibleContext & \cite{feng2025struphantom} \\ \cline{3-8} 
 &  & \AtkDirect & \VictimApp:Defense & \GeneralProperty:Defend Bypass & \BlackBox:Defender Feedback & \InVisible:\InVisibleSemantics & \cite{liu2025aegis} \\ \cline{3-8} 
 &  & \AtkIndirect & \VictimAgent:General & \InteProperty:Goal Hijack & \WhiteBox:Logits & \InVisible:\InVisibleContext & \cite{zhan2025adaptive} \\ \cline{3-8} 
 &  & \AtkIndirect & \VictimApp:Search & \InteProperty:Rank Tamper & \BlackBox:Rankings & \InVisible:\InVisibleSemantics & \cite{nestaas2024adversarial} \\ \cline{2-8} 
 & \multirow{2}{*}{Sampling} & \AtkDirect & \VictimLLM:Base & \InteProperty:Goal Hijack & \BlackBox:Reward Signal & \InVisible:\InVisibleSemantics & \cite{wen2025rl} \\ \cline{3-8} 
 &  & \AtkDirect & \VictimLLM:Base & \InteProperty:Goal Hijack & \BlackBox:Surrogate Activations & \InVisible:\InVisibleSemantics & \cite{li2025transferable} \\ \hline\hline
\multirow{23}{*}{Heuristic} & \multirow{16}{*}{Manual Template} & \AtkDirect & \VictimAgent:Memory & \InteProperty:Action Hijack & \BlackBox:Memory Output & \InVisible:\InVisibleContext & \cite{zhang2024towards} \\ \cline{3-8} 
 &  & \AtkIndirect & \VictimAgent:Multi & \InteProperty:Agent Hijack & \BlackBox:Consensus State & \InVisible:\InVisibleContext & \cite{cui2025mad} \\ \cline{3-8} 
 &  & \AtkIndirect & \VictimAgent:Multi & \InteProperty:Agent Hijack & \BlackBox:Message Passing & \InVisible:\InVisibleContext & \cite{lee2024prompt} \\ \cline{3-8} 
 &  & \AtkIndirect & \VictimAgent:General & \ConfProperty:Data Leakage & \BlackBox:Tool Output & \Visible & \cite{alizadeh2025simple} \\ \cline{3-8} 
 &  & \AtkDirect/\AtkIndirect & \VictimLLM:Finance & \InteProperty:Goal Hijack & \BlackBox:Output Text & \Visible & \cite{chang2025breaking} \\ \cline{3-8} 
 &  & \AtkIndirect & \VictimAgent:Hacker & \InteProperty:Action Hijack & \BlackBox:Logs & \InVisible:\InVisibleContext & \cite{pasquini2024hacking} \\ \cline{3-8} 
 &  & \AtkDirect/\AtkIndirect & \VictimLLM:Medical & \InteProperty:Misinformation & \BlackBox:Text Output & \Visible & \cite{clusmann2024prompt} \\ \cline{3-8} 
 &  & \AtkDirect & \VictimLLM:CoT & \AvaiProperty:CoT DoS & \BlackBox:Reasoning Trace & \Visible & \cite{xu2024preemptive} \\ \cline{3-8} 
 &  & \AtkDirect & \VictimApp:Translator & \InteProperty:Task Hijack & \BlackBox:Translation & \Visible & \cite{sun2024scaling} \\ \cline{3-8} 
 &  & \AtkIndirect & \VictimAgent:Coding & \InteProperty:Action Hijack & \BlackBox:Execution & \InVisible:\InVisibleVision & \cite{liu2025your} \\ \cline{3-8} 
 &  & \AtkIndirect & \VictimApp:Product & \ConfProperty:Data Leakage & \BlackBox:Exfiltration Log & \InVisible:\InVisibleContext & \cite{reddy2025echoleak} \\ \cline{3-8} 
 &  & \AtkIndirect & \VictimAgent:Hacker & \InteProperty:Action Hijack & \BlackBox:Shell Access & \InVisible:\InVisibleEncoding & \cite{mayoral2025cybersecurity} \\ \cline{3-8} 
 &  & \AtkIndirect & \VictimApp:Reviewer & \InteProperty:Score Tamper & \BlackBox:Review Output & \InVisible:\InVisibleVision & \cite{zhu2025your} \\ \cline{3-8} 
 &  & \AtkDirect & \VictimApp:Evaluator & \InteProperty:Score Tamper & \BlackBox:Score Output & \InVisible:\InVisibleEncoding & \cite{maloyan2025adversarial} \\ \cline{3-8} 
 &  & \AtkSupply & \VictimLLM:Base & \GeneralProperty:Defend Bypass & \WhiteBox:Training Data & \Visible & \cite{shao2024enhancing} \\ \cline{3-8} 
 &  & \AtkSupply & \VictimLLM:Base & \GeneralProperty:Defend Bypass & \WhiteBox:Training Access & \Visible & \cite{chen2025backdoor} \\ \cline{2-8} 
 & \multirow{3}{*}{LLM Generation} & \AtkDirect & \VictimLLM:CoT & \AvaiProperty:CoT DoS & \BlackBox:Output Length & \InVisible:\InVisibleSemantics & \cite{cui2025token} \\ \cline{3-8} 
 &  & \AtkIndirect & \VictimLLM:RAG & \ConfProperty:Data Leakage & \BlackBox:Leak Success & \InVisible:\InVisibleContext & \cite{cui2025vortexpia} \\ \cline{3-8} 
 &  & \AtkDirect & \VictimApp:Evaluator & \GeneralProperty:Defend Bypass & \BlackBox:Score Consistency & \Visible & \cite{liu2025counterfactual} \\ \cline{2-8} 
 & \multirow{4}{*}{Structural Encoding} & \AtkIndirect & \VictimApp:Reviewer & \InteProperty:Score Tamper & \BlackBox:Review Score & \InVisible:\InVisibleVision & \cite{collu2025publish} \\ \cline{3-8} 
 &  & \AtkIndirect & \VictimApp:Reviewer & \InteProperty:Score Tamper & \BlackBox:Grading Score & \Visible & \cite{guo2025too} \\ \cline{3-8} 
 &  & \AtkIndirect & \VictimAgent:Browser & \ConfProperty:Data Leakage & \BlackBox:Web Request & \InVisible:\InVisibleEncoding & \cite{rall2025exploiting} \\ \cline{3-8} 
 &  & \AtkIndirect & \VictimApp:Reviewer & \InteProperty:Score Tamper & \BlackBox:Review Score & \InVisible:\InVisibleContext & \cite{keuper2025prompt} \\ \hline\hline
\end{tabular}
\begin{tablenotes}[para, flushleft]
\footnotesize
\item \textbf{(1) For the attack surface}:
    \AtkDirect: direct prompt injection;
    \AtkIndirect: indirect prompt injection;
    \AtkSupply: prompt injection from supply chain.
\textbf{(2) For the victim}:
    \VictimLLM: LLM;
    \VictimApp: LLM-integrated application;
    \VictimAgent: LLM agent.
\textbf{(3) For the access}:
    \BlackBox: black-box;
    \WhiteBox: white-box.
\textbf{(4) For the perceptibility}:
    \Visible: visible payloads;
    \InVisible: invisible payloads;
    \InVisibleVision: invisible in vision;
    \InVisibleContext: invisible via merging in context;
    \InVisibleEncoding: invisible via encoding;
    \InVisibleSemantics: invisible via adversarial optimization of semantics.
\end{tablenotes}
\end{threeparttable}
\end{table*}



\subsection{Attack Payload Generation}\label{sec:attack:method}



We systematize the collected 37 prompt injection attack papers based on their payload generation methodologies, as summarized in Table~\ref{tab:attack-taxonomy}. 
We categorize these methods into two paradigms: \textit{heuristic-based} approaches, which rely on manual design or semantic exploitation strategies (23 out of 37 works) to generate payloads, and \textit{optimization-based} approaches, which employ automated algorithms to search for optimal payloads (14 out of 37 works). 

\sssec{Heuristic}
Heuristic-based attacks exploit the intrinsic \textit{instruction-following bias} of LLMs, typically treating the target agent as a black box. 
We categorize these works (\textit{23 out of 37 papers}) into 3 subtypes: 

\underline{(1) Manual template}. As the most prevalent category (\textit{16 papers}), these attacks involve manually constructing adversarial prompts that exploit priority conflicts in the attention mechanism, where the model prioritizes recent or authoritative-sounding instructions over system prompts. 
While early techniques focused on direct overrides, recent works demonstrate that such templates have evolved into stealthy IPI embedded within agent memory or log files to trigger action hijacking~\cite{zhang2024towards, pasquini2024hacking}. 
Furthermore, this vector extends to the supply chain, where attackers poison training datasets with backdoor triggers to permanently compromise model alignment~\cite{shao2024enhancing, chen2025backdoor}.

\underline{(2) LLM generation}. 
To address the inefficiency of manual crafting, researchers employ ``LLM-against-LLM'' frameworks (\textit{3 papers}). 
Unlike optimization-based methods that iteratively search for payloads, these approaches leverage the generative capabilities of a Red-team LLM to generate adversarial instructions based on heuristic rules ~\cite{cui2025token, cui2025vortexpia}. 

\underline{(3) Structural encoding}. 
These attacks (\textit{4 papers}) target the cognitive gap between the model's tokenizer and its semantic processing. 
Instead of relying on natural language, adversaries encode payloads into non-semantic formats, such as Base64, ASCII art, or structured file layouts, that bypass semantic safety filters while remaining executable. 
Recent studies validate that such structural injections can manipulate logic in PDF parsing or web search tools, highlighting the insufficiency of semantic-only defenses~\cite{collu2025publish, keuper2025prompt, rall2025exploiting}.

\sssec{Optimization}
Optimization-based attacks automate the search for adversarial suffixes or token combinations that maximize the likelihood of malicious generation. 
These methods (\textit{14 out of 37 papers}) are categorized by their access requirements into gradient-based (white-box) and genetic/sampling-based (black-box) approaches.

\underline{(1) Gradient}. 
Requiring white-box access to model weights (\textit{6 papers}), these methods compute the gradient of the loss function with respect to input tokens. Inspired by Greedy Coordinate Gradient (GCG) attacks, they iteratively update the payload to minimize model resistance. 
Applications include generating universal adversarial suffixes to bypass perplexity filters, fingerprinting LLMs via injection response patterns, and crafting neural execution triggers to evade sanitization layers~\cite{liu2024automatic, hu2025fingerprinting, pasquini2024neural}.

\underline{(2) Genetic and (3) sampling}. To operate under black-box constraints (\textit{8 papers}), researchers utilize evolutionary algorithms or sampling techniques driven by query feedback. 
Genetic approaches evolve a population of prompts to bypass distributional detectors or manipulate tabular agents~\cite{zhang2024goal, feng2025struphantom}. 
Alternatively, Reinforcement Learning (RL) and MCMC sampling frameworks model the attack as a reward maximization problem, generating transferable injections that bypass instruction hierarchies without direct gradient access~\cite{wen2025rl, li2025transferable}.

\takeaways{
\textbf{Misalignment between attack payload generation and defense evaluation.}
Our analysis highlights a critical ``evaluation gap'': while optimization-based attacks now constitute \textit{14 out of 37 works}, existing defenses and benchmarks continue to evaluate security primarily against heuristic templates. 
This misalignment creates a ``false sense of security'', where defenses appear robust against manual patterns but remain untested against automated-generated payloads. 

}

\subsection{Attack Paradigms}\label{sec:attack:dimension}

In this section, we discuss the core attack paradigm across all attacks, including the threat models (surfaces, victims, and goals), the attacker capabilities, and the visibility of payloads.

\sssec{Threat model}.
Mapping the threat landscape to the execution loop in \S\ref{sec:preliminary:lifecycle}, we categorize attacks into three vectors: \textit{Direct Prompt Injection (DPI)} \AtkDirect (malicious user inputs), \textit{Indirect Prompt Injection (IPI)} \AtkIndirect (adversarial instructions embedded in external resources), and \textit{Supply-chain Prompt Injection (SPI)} \AtkSupply (payloads within RAG or training datasets). 
Our analysis reveals a significant expansion in the attack surface: while foundational studies address \AtkDirect DPI (\textit{17 works}), the majority of agent-specific research now prioritizes \AtkIndirect IPI (\textit{20 works}).
At the same time, attack victims have gradually shifted from \VictimLLM base LLM to researching specific \VictimApp LLM applications and \VictimAgent LLM agents.
Consequently, attacker goals have shifted from safety violations to \InteProperty integrity compromise (\textit{30 out of 37 works}), manifesting primarily as action or goal hijacking. 
Notably, \ConfProperty confidentiality attacks (\textit{4 works})~\cite{alizadeh2025simple, reddy2025echoleak, cui2025vortexpia, rall2025exploiting} increasingly converge with hijacking for data exfiltration, while \AvaiProperty availability vectors (\textit{2 works})~\cite{xu2024preemptive, cui2025token} specifically target reasoning mechanisms (e.g., CoT DoS).

\sssec{Attacker capabilities}.
We categorize capabilities by system access, where a minority of studies (\textit{9 works}) assume \WhiteBox white-box access to model parameters, leveraging \WhiteBox: gradients for adversarial suffixes~\cite{liu2024automatic, pasquini2024neural, zhang2024hijackrag}, \WhiteBox: logits for fingerprinting~\cite{hu2025fingerprinting, shi2024optimization}, or \WhiteBox: training data for poisoning~\cite{shao2024enhancing, chen2025backdoor}. 
Conversely, the majority (\textit{28 out of 37}) operate under \BlackBox black-box constraints, optimizing against visible \BlackBox: text/score outputs~\cite{maloyan2025adversarial, sun2024scaling} or restricted \BlackBox: success boolean signals~\cite{yu2024promptfuzz, cui2025vortexpia}. 
Notably, agentic systems introduce environmental side-channels, enabling state inference via \BlackBox: logs~\cite{pasquini2024hacking}, \BlackBox: memory output~\cite{zhang2024towards}, or \BlackBox: execution effects~\cite{liu2025your, mayoral2025cybersecurity} without direct model access.

\sssec{Attack visibility}
We classify payloads into \Visible visible and \InVisible invisible categories, identifying a paradigm shift where the majority of research (\textit{27 out of 37 works}) focuses on invisibility to evade detection. While early heuristics employed \Visible visible payloads with explicit natural language triggers (\textit{10 works}), adversaries have pivoted to stealthier vectors: semantics invisibility (\textit{11 works}) utilizes optimization to craft non-meaningful suffixes~\cite{liu2024automatic}; context invisibility (\textit{10 works}) conceals payloads within massive context windows via poisoned sources like RAG~\cite{zhang2024hijackrag} or logs~\cite{pasquini2024hacking}; encoding invisibility (\textit{3 works}) exploits parsing gaps via non-standard formats (e.g., Base64)~\cite{rall2025exploiting}; and vision invisibility (\textit{3 works}) embeds imperceptible instructions into visual inputs transparent to humans but legible to agents~\cite{zhu2025your}.

\takeaways{
\textbf{The paradigm shift to more practical attacks.}
The threat landscape has transitioned from theoretical safety violations to practical integrity compromises, where adversaries prioritize high-stakes action hijacking over simple toxic generation. 
This evolution is characterized by a migration from direct, visible overrides to stealthy, environment-driven vectors, specifically IPI and invisible optimization, that exploit the agent's context processing rather than relying on white-box model access.
}

\begin{figure}[!htbp]
    \centering
    \includegraphics[width=\linewidth]{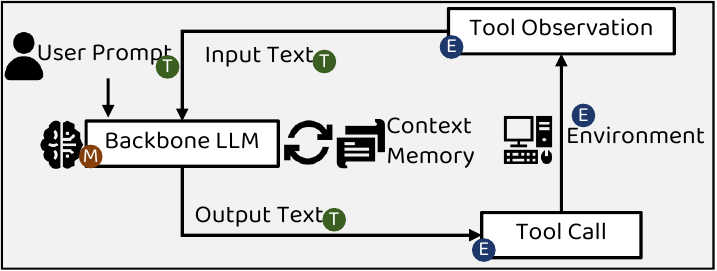}
    \caption{
We map the taxonomy of defenses to 3 different levels:
(1) \circledGreen{T} Text-level: Stateless defenses focusing on the backbone LLMs' input and output;
(2) \circledOrange{M} Model-level: Internal defenses focusing on the model parameters or inference internal representation (IR);
(3) \circledBlue{E} Execution-level: Stateful defenses focusing on the consequences and causality of actions within the environments.
    }
    \label{fig:defense_tax}
\end{figure}

\section{Taxonomy of Defenses}\label{sec:defense_survey}

\begin{table*}[!t]
\footnotesize
\centering
\begin{threeparttable}
\caption{Systematization of prompt injection defenses, categorized into 3 levels as illustrated in Fig. \ref{fig:defense_tax}.}
\label{tab:defense_taxonomy}
\begin{tabular}{|cll|ccc|cc|cccc|cc|}
\hline\hline
\multicolumn{3}{|c|}{Category} & \multicolumn{3}{c|}{Capability \S\ref{sec:defense:perf}} & \multicolumn{2}{c|}{Explain. \S\ref{sec:defense:explain}} & \multicolumn{4}{c|}{Cost \S\ref{sec:defense:cost}} & \multicolumn{2}{c|}{Paper} \\ \hline
\multicolumn{1}{|c|}{Level} & \multicolumn{1}{l|}{Strategy} & Method & \multicolumn{1}{c|}{Stage} & \multicolumn{1}{c|}{Granular} & Attr. & \multicolumn{1}{c|}{Method} & Rely & \multicolumn{1}{c|}{Compu.} & \multicolumn{1}{c|}{Compa.} & \multicolumn{1}{c|}{Auto.} & Uti. & \multicolumn{1}{c|}{Ref.} & Code \\ \hline\hline
\multicolumn{1}{|c|}{\multirow{18}{*}{\begin{tabular}[c]{@{}c@{}}\S\ref{sec:defense_survey:text}\\ Text\end{tabular}}} & \multicolumn{1}{l|}{\multirow{11}{*}{\begin{tabular}[c]{@{}l@{}}Detection\\ Filter\end{tabular}}} & \multirow{7}{*}{LLM} & \multicolumn{1}{c|}{\TextInput} & \multicolumn{1}{c|}{Message} & \InteProperty & \multicolumn{1}{c|}{--} & -- & \multicolumn{1}{c|}{Model} & \multicolumn{1}{c|}{\LowCompa} & \multicolumn{1}{c|}{\FullAuto} & \UtilityStatic & \multicolumn{1}{c|}{\cite{jacob2024promptshield}} & \ding{55} \\ \cline{4-14} 
\multicolumn{1}{|c|}{} & \multicolumn{1}{l|}{} &  & \multicolumn{1}{c|}{\TextInput \TextOutput} & \multicolumn{1}{c|}{Message} & \InteProperty & \multicolumn{1}{c|}{Causal} & \ExplainLLM & \multicolumn{1}{c|}{LLM} & \multicolumn{1}{c|}{\LowCompa} & \multicolumn{1}{c|}{\FullAuto} & \UtilityHuman & \multicolumn{1}{c|}{\cite{pan2025prompt}} & \ding{55} \\ \cline{4-14} 
\multicolumn{1}{|c|}{} & \multicolumn{1}{l|}{} &  & \multicolumn{1}{c|}{\TextInput} & \multicolumn{1}{c|}{Inst/Data} & \InteProperty \ConfProperty & \multicolumn{1}{c|}{Causal} & \ExplainSem & \multicolumn{1}{c|}{LLM} & \multicolumn{1}{c|}{\LowCompa} & \multicolumn{1}{c|}{\FullAuto} & \UtilityStatic & \multicolumn{1}{c|}{\cite{shi2025promptarmor}} & \ding{55} \\ \cline{4-14} 
\multicolumn{1}{|c|}{} & \multicolumn{1}{l|}{} &  & \multicolumn{1}{c|}{\TextOutput} & \multicolumn{1}{c|}{Context} & \InteProperty & \multicolumn{1}{c|}{Causal} & \ExplainSem & \multicolumn{1}{c|}{LLM} & \multicolumn{1}{c|}{\MedCompa} & \multicolumn{1}{c|}{\FullAuto} & \UtilityStatic & \multicolumn{1}{c|}{\cite{liu2025secinfer}} & \ding{55} \\ \cline{4-14} 
\multicolumn{1}{|c|}{} & \multicolumn{1}{l|}{} &  & \multicolumn{1}{c|}{\TextInput \SystemPrompt} & \multicolumn{1}{c|}{Message} & \InteProperty & \multicolumn{1}{c|}{Policy} & \ExplainHuman & \multicolumn{1}{c|}{--} & \multicolumn{1}{c|}{\LowCompa} & \multicolumn{1}{c|}{\SemiAuto} & \UtilityStatic & \multicolumn{1}{c|}{\cite{pawelek2025llmz+}} & \ding{55} \\ \cline{4-14} 
\multicolumn{1}{|c|}{} & \multicolumn{1}{l|}{} &  & \multicolumn{1}{c|}{\TextInput} & \multicolumn{1}{c|}{Segment} & \InteProperty & \multicolumn{1}{c|}{Causal} & \ExplainSem & \multicolumn{1}{c|}{LLM} & \multicolumn{1}{c|}{\LowCompa} & \multicolumn{1}{c|}{\FullAuto} & \UtilityStatic & \multicolumn{1}{c|}{\cite{jia2025promptlocate}} & {\ul \github{https://github.com/liu00222/Open-Prompt-Injection}} \\ \cline{4-14} 
\multicolumn{1}{|c|}{} & \multicolumn{1}{l|}{} &  & \multicolumn{1}{c|}{\ToolObs} & \multicolumn{1}{c|}{Segment} & \InteProperty & \multicolumn{1}{c|}{--} & -- & \multicolumn{1}{c|}{LLM} & \multicolumn{1}{c|}{\MedCompa} & \multicolumn{1}{c|}{\FullAuto} & \UtilityStatic & \multicolumn{1}{c|}{\cite{kerboua2025focusagent}} & \ding{55} \\ \cline{3-14} 
\multicolumn{1}{|c|}{} & \multicolumn{1}{l|}{} & \multirow{4}{*}{Non-LLM} & \multicolumn{1}{c|}{\TextInput} & \multicolumn{1}{c|}{Message} & \InteProperty & \multicolumn{1}{c|}{--} & -- & \multicolumn{1}{c|}{Model} & \multicolumn{1}{c|}{\LowCompa} & \multicolumn{1}{c|}{\FullAuto} & \UtilityStatic & \multicolumn{1}{c|}{\cite{rahman2024applying}} & \ding{55} \\ \cline{4-14} 
\multicolumn{1}{|c|}{} & \multicolumn{1}{l|}{} &  & \multicolumn{1}{c|}{\TextInput} & \multicolumn{1}{c|}{Message} & \InteProperty & \multicolumn{1}{c|}{--} & -- & \multicolumn{1}{c|}{Model} & \multicolumn{1}{c|}{\LowCompa} & \multicolumn{1}{c|}{\FullAuto} & \UtilityStatic & \multicolumn{1}{c|}{\cite{li2025piguard}} & {\ul \github{https://github.com/leolee99/PIGuard}} \\ \cline{4-14} 
\multicolumn{1}{|c|}{} & \multicolumn{1}{l|}{} &  & \multicolumn{1}{c|}{\TextInput} & \multicolumn{1}{c|}{Segment} & \InteProperty & \multicolumn{1}{c|}{--} & -- & \multicolumn{1}{c|}{Model} & \multicolumn{1}{c|}{\LowCompa} & \multicolumn{1}{c|}{\FullAuto} & \UtilityStatic & \multicolumn{1}{c|}{\cite{chen2025can}} & {\ul \github{https://github.com/LukeChen-go/indirect-pia-detection}} \\ \cline{4-14} 
\multicolumn{1}{|c|}{} & \multicolumn{1}{l|}{} &  & \multicolumn{1}{c|}{\TextInput \ToolObs} & \multicolumn{1}{c|}{Token} & \InteProperty & \multicolumn{1}{c|}{--} & -- & \multicolumn{1}{c|}{Model} & \multicolumn{1}{c|}{\LowCompa} & \multicolumn{1}{c|}{\FullAuto} & \UtilityStatic & \multicolumn{1}{c|}{\cite{das2025commandsans}} & \ding{55} \\ \cline{2-14} 
\multicolumn{1}{|c|}{} & \multicolumn{1}{l|}{\multirow{7}{*}{\begin{tabular}[c]{@{}l@{}}Prompt\\ Enhance\end{tabular}}} & \multirow{3}{*}{\begin{tabular}[c]{@{}l@{}}I/O\\ Separate\end{tabular}} & \multicolumn{1}{c|}{\TextInput \SystemPrompt} & \multicolumn{1}{c|}{Inst/Data} & \InteProperty \ConfProperty & \multicolumn{1}{c|}{--} & -- & \multicolumn{1}{c|}{--} & \multicolumn{1}{c|}{\MedCompa} & \multicolumn{1}{c|}{\FullAuto} & \UtilityStatic & \multicolumn{1}{c|}{\cite{hines2024defending}} & \ding{55} \\ \cline{4-14} 
\multicolumn{1}{|c|}{} & \multicolumn{1}{l|}{} &  & \multicolumn{1}{c|}{\SystemPrompt \TextOutput} & \multicolumn{1}{c|}{Inst/Data} & \InteProperty & \multicolumn{1}{c|}{--} & -- & \multicolumn{1}{c|}{--} & \multicolumn{1}{c|}{\MedCompa} & \multicolumn{1}{c|}{\FullAuto} & \UtilityStatic & \multicolumn{1}{c|}{\cite{wang2024fath}} & {\ul \github{https://github.com/Jayfeather1024/FATH}} \\ \cline{4-14} 
\multicolumn{1}{|c|}{} & \multicolumn{1}{l|}{} &  & \multicolumn{1}{c|}{\SystemPrompt \TextOutput} & \multicolumn{1}{c|}{Inst/Data} & \InteProperty & \multicolumn{1}{c|}{Causal} & \ExplainLLM & \multicolumn{1}{c|}{--} & \multicolumn{1}{c|}{\MedCompa} & \multicolumn{1}{c|}{\FullAuto} & \UtilityStatic & \multicolumn{1}{c|}{\cite{chen2025robustness}} & \ding{55} \\ \cline{3-14} 
\multicolumn{1}{|c|}{} & \multicolumn{1}{l|}{} & \multirow{2}{*}{\begin{tabular}[c]{@{}l@{}}Negative\\ Prompts\end{tabular}} & \multicolumn{1}{c|}{\TextInput} & \multicolumn{1}{c|}{Inst/Data} & \InteProperty & \multicolumn{1}{c|}{--} & -- & \multicolumn{1}{c|}{--} & \multicolumn{1}{c|}{\MedCompa} & \multicolumn{1}{c|}{\FullAuto} & \UtilityStatic & \multicolumn{1}{c|}{\cite{chen2025defense}} & {\ul \github{https://github.com/LukeChen-go/pia-defense-by-attack}} \\ \cline{4-14} 
\multicolumn{1}{|c|}{} & \multicolumn{1}{l|}{} &  & \multicolumn{1}{c|}{\SystemPrompt} & \multicolumn{1}{c|}{Context} & \InteProperty & \multicolumn{1}{c|}{--} & -- & \multicolumn{1}{c|}{--} & \multicolumn{1}{c|}{\MedCompa} & \multicolumn{1}{c|}{\FullAuto} & \UtilityStatic & \multicolumn{1}{c|}{\cite{chen2025defending}} & {\ul \github{https://github.com/Sizhe-Chen/DefensiveToken}} \\ \cline{3-14} 
\multicolumn{1}{|c|}{} & \multicolumn{1}{l|}{} & \multirow{2}{*}{Rewrite} & \multicolumn{1}{c|}{\TextInput \SystemPrompt} & \multicolumn{1}{c|}{Message} & \InteProperty & \multicolumn{1}{c|}{Policy} & \ExplainHuman & \multicolumn{1}{c|}{--} & \multicolumn{1}{c|}{\LowCompa} & \multicolumn{1}{c|}{\SemiAuto} & \UtilityStatic & \multicolumn{1}{c|}{\cite{alharthi2025call}} & \ding{55} \\ \cline{4-14} 
\multicolumn{1}{|c|}{} & \multicolumn{1}{l|}{} &  & \multicolumn{1}{c|}{\SystemPrompt} & \multicolumn{1}{c|}{Context} & \InteProperty & \multicolumn{1}{c|}{--} & -- & \multicolumn{1}{c|}{--} & \multicolumn{1}{c|}{\MedCompa} & \multicolumn{1}{c|}{\FullAuto} & \UtilityStatic & \multicolumn{1}{c|}{\cite{wang2025protect}} & {\ul \github{https://github.com/zhilongwang/LLMAgentProtector}} \\ \hline\hline
\multicolumn{1}{|c|}{\multirow{7}{*}{\begin{tabular}[c]{@{}c@{}}\S\ref{sec:defense_survey:model}\\ Model\end{tabular}}} & \multicolumn{1}{l|}{\multirow{4}{*}{\begin{tabular}[c]{@{}l@{}}Model\\ Align\end{tabular}}} & Task & \multicolumn{1}{c|}{\ModelParam} & \multicolumn{1}{c|}{Context} & \InteProperty & \multicolumn{1}{c|}{--} & -- & \multicolumn{1}{c|}{Finetune} & \multicolumn{1}{c|}{\MedCompa} & \multicolumn{1}{c|}{\FullAuto} & \UtilityStatic & \multicolumn{1}{c|}{\cite{piet2024jatmo}} & {\ul \github{https://github.com/wagner-group/prompt-injection-defense}} \\ \cline{3-14} 
\multicolumn{1}{|c|}{} & \multicolumn{1}{l|}{} & Preference & \multicolumn{1}{c|}{\ModelParam} & \multicolumn{1}{c|}{Inst/Data} & \InteProperty & \multicolumn{1}{c|}{--} & -- & \multicolumn{1}{c|}{Finetune} & \multicolumn{1}{c|}{\MedCompa} & \multicolumn{1}{c|}{\FullAuto} & \UtilityStatic & \multicolumn{1}{c|}{\cite{chen2025secalign}} & {\ul \github{https://github.com/facebookresearch/SecAlign}} \\ \cline{3-14} 
\multicolumn{1}{|c|}{} & \multicolumn{1}{l|}{} & \multirow{2}{*}{Format} & \multicolumn{1}{c|}{\TextInput \ModelParam} & \multicolumn{1}{c|}{Context} & \InteProperty & \multicolumn{1}{c|}{--} & -- & \multicolumn{1}{c|}{Finetune} & \multicolumn{1}{c|}{\MedCompa} & \multicolumn{1}{c|}{\FullAuto} & \UtilityStatic & \multicolumn{1}{c|}{\cite{chen2025struq}} & {\ul \github{https://github.com/Sizhe-Chen/StruQ}} \\ \cline{4-14} 
\multicolumn{1}{|c|}{} & \multicolumn{1}{l|}{} &  & \multicolumn{1}{c|}{\SystemPrompt} & \multicolumn{1}{c|}{Message} & \InteProperty & \multicolumn{1}{c|}{Causal} & \ExplainIR & \multicolumn{1}{c|}{Finetune} & \multicolumn{1}{c|}{\MedCompa} & \multicolumn{1}{c|}{\FullAuto} & \UtilityStatic & \multicolumn{1}{c|}{\cite{wang2025protect}} & {\ul \github{https://github.com/zhilongwang/LLMAgentProtector}} \\ \cline{2-14} 
\multicolumn{1}{|c|}{} & \multicolumn{1}{l|}{\multirow{3}{*}{\begin{tabular}[c]{@{}l@{}}Model IR\\ Intervene\end{tabular}}} & \multirow{3}{*}{\begin{tabular}[c]{@{}l@{}}IR-based\\ Detector\end{tabular}} & \multicolumn{1}{c|}{\ModelIR} & \multicolumn{1}{c|}{Context} & \InteProperty & \multicolumn{1}{c|}{Causal} & \ExplainIR & \multicolumn{1}{c|}{Probe} & \multicolumn{1}{c|}{\LowCompa} & \multicolumn{1}{c|}{\FullAuto} & \UtilityStatic & \multicolumn{1}{c|}{\cite{wen2025defending}} & {\ul \github{https://github.com/MYVAE/Instruction-detection}} \\ \cline{4-14} 
\multicolumn{1}{|c|}{} & \multicolumn{1}{l|}{} &  & \multicolumn{1}{c|}{\ModelIR} & \multicolumn{1}{c|}{Segment} & \InteProperty & \multicolumn{1}{c|}{Causal} & \ExplainIR & \multicolumn{1}{c|}{Probe} & \multicolumn{1}{c|}{\LowCompa} & \multicolumn{1}{c|}{\FullAuto} & \UtilityStatic & \multicolumn{1}{c|}{\cite{hung2025attention}} & \ding{55} \\ \cline{4-14} 
\multicolumn{1}{|c|}{} & \multicolumn{1}{l|}{} &  & \multicolumn{1}{c|}{\ModelIR} & \multicolumn{1}{c|}{Message} & \InteProperty & \multicolumn{1}{c|}{Causal} & \ExplainIR & \multicolumn{1}{c|}{Probe} & \multicolumn{1}{c|}{\LowCompa} & \multicolumn{1}{c|}{\FullAuto} & \UtilityStatic & \multicolumn{1}{c|}{\cite{zou2025pishield}} & {\ul \github{https://github.com/weizou52/PIShield}} \\ \hline\hline
\multicolumn{1}{|c|}{\multirow{16}{*}{\begin{tabular}[c]{@{}c@{}}\S\ref{sec:defense_survey:execution}\\ Execution\end{tabular}}} & \multicolumn{1}{l|}{\multirow{2}{*}{\begin{tabular}[c]{@{}l@{}}Task\\ Align\end{tabular}}} & Goal Align & \multicolumn{1}{c|}{\ToolCall \TextOutput} & \multicolumn{1}{c|}{Inst/Data} & \InteProperty & \multicolumn{1}{c|}{Causal} & \ExplainLLM & \multicolumn{1}{c|}{LLM} & \multicolumn{1}{c|}{\LowCompa} & \multicolumn{1}{c|}{\FullAuto} & \UtilityDynamic & \multicolumn{1}{c|}{\cite{jia2025task}} & \ding{55} \\ \cline{3-14} 
\multicolumn{1}{|c|}{} & \multicolumn{1}{l|}{} & Inject Align & \multicolumn{1}{c|}{\ToolCall \TextInput} & \multicolumn{1}{c|}{Context} & \InteProperty \ConfProperty & \multicolumn{1}{c|}{Causal} & \ExplainLLM & \multicolumn{1}{c|}{LLM} & \multicolumn{1}{c|}{\LowCompa} & \multicolumn{1}{c|}{\FullAuto} & \UtilityDynamic & \multicolumn{1}{c|}{\cite{zhu2025melon}} & {\ul \github{https://github.com/kaijiezhu11/MELON}} \\ \cline{2-14} 
\multicolumn{1}{|c|}{} & \multicolumn{1}{l|}{\multirow{10}{*}{\begin{tabular}[c]{@{}l@{}}Access\\ Control\end{tabular}}} & \multirow{7}{*}{\begin{tabular}[c]{@{}l@{}}Flow\\ Control\end{tabular}} & \multicolumn{1}{c|}{\AgentArch \Environment} & \multicolumn{1}{c|}{Inst/Data} & \InteProperty \ConfProperty & \multicolumn{1}{c|}{Causal} & \ExplainLLM\ExplainHuman & \multicolumn{1}{c|}{LLM} & \multicolumn{1}{c|}{\LowCompa} & \multicolumn{1}{c|}{\SemiAuto} & \UtilityDynamic & \multicolumn{1}{c|}{\cite{costa2025securing}} & {\ul \github{https://github.com/microsoft/fides}} \\ \cline{4-14} 
\multicolumn{1}{|c|}{} & \multicolumn{1}{l|}{} &  & \multicolumn{1}{c|}{\AgentArch \Environment} & \multicolumn{1}{c|}{Inst/Data} & \InteProperty \ConfProperty & \multicolumn{1}{c|}{--} & -- & \multicolumn{1}{c|}{LLM} & \multicolumn{1}{c|}{\HighCompa} & \multicolumn{1}{c|}{\SemiAuto} & \UtilityHuman & \multicolumn{1}{c|}{\cite{li2025safeflow}} & \ding{55} \\ \cline{4-14} 
\multicolumn{1}{|c|}{} & \multicolumn{1}{l|}{} &  & \multicolumn{1}{c|}{\Environment} & \multicolumn{1}{c|}{Inst/Data} & \InteProperty \ConfProperty & \multicolumn{1}{c|}{Causal} & \ExplainLLM\ExplainHuman & \multicolumn{1}{c|}{Other} & \multicolumn{1}{c|}{\LowCompa} & \multicolumn{1}{c|}{\SemiAuto} & \UtilityDynamic & \multicolumn{1}{c|}{\cite{siddiqui2024permissive}} & \ding{55} \\ \cline{4-14} 
\multicolumn{1}{|c|}{} & \multicolumn{1}{l|}{} &  & \multicolumn{1}{c|}{\AgentArch \Environment} & \multicolumn{1}{c|}{Inst/Data} & \InteProperty \ConfProperty & \multicolumn{1}{c|}{--} & -- & \multicolumn{1}{c|}{Other} & \multicolumn{1}{c|}{\HighCompa} & \multicolumn{1}{c|}{\FullAuto} & \UtilityStatic & \multicolumn{1}{c|}{\cite{wu2024system}} & {\ul \github{https://github.com/fzwark/Secure_LLM_System}} \\ \cline{4-14} 
\multicolumn{1}{|c|}{} & \multicolumn{1}{l|}{} &  & \multicolumn{1}{c|}{\ModelIR \AgentArch} & \multicolumn{1}{c|}{Inst/Data} & \InteProperty \ConfProperty & \multicolumn{1}{c|}{Causal} & \ExplainLLM\ExplainHuman & \multicolumn{1}{c|}{LLM} & \multicolumn{1}{c|}{\LowCompa} & \multicolumn{1}{c|}{\SemiAuto} & \UtilityHuman & \multicolumn{1}{c|}{\cite{zhong2025rtbas}} & \ding{55} \\ \cline{4-14} 
\multicolumn{1}{|c|}{} & \multicolumn{1}{l|}{} &  & \multicolumn{1}{c|}{\AgentArch \Environment} & \multicolumn{1}{c|}{Inst/Data} & \InteProperty \ConfProperty & \multicolumn{1}{c|}{Causal} & \ExplainLLM\ExplainHuman & \multicolumn{1}{c|}{Code} & \multicolumn{1}{c|}{\HighCompa} & \multicolumn{1}{c|}{\SemiAuto} & \UtilityStatic & \multicolumn{1}{c|}{\cite{debenedetti2025defeating}} & {\ul \github{https://github.com/google-research/camel-prompt-injection}} \\ \cline{4-14} 
\multicolumn{1}{|c|}{} & \multicolumn{1}{l|}{} &  & \multicolumn{1}{c|}{\AgentArch \ToolCall} & \multicolumn{1}{c|}{Inst/Data} & \InteProperty \ConfProperty & \multicolumn{1}{c|}{Causal} & \ExplainLLM & \multicolumn{1}{c|}{LLM} & \multicolumn{1}{c|}{\LowCompa} & \multicolumn{1}{c|}{\FullAuto} & \UtilityStatic & \multicolumn{1}{c|}{\cite{wang2025agentarmor}} & \ding{55} \\ \cline{3-14} 
\multicolumn{1}{|c|}{} & \multicolumn{1}{l|}{} & \multirow{3}{*}{\begin{tabular}[c]{@{}l@{}}Spec\\ Control\end{tabular}} & \multicolumn{1}{c|}{\AgentArch} & \multicolumn{1}{c|}{Message} & \InteProperty \ConfProperty & \multicolumn{1}{c|}{Policy} & \ExplainLLM & \multicolumn{1}{c|}{LLM} & \multicolumn{1}{c|}{\LowCompa} & \multicolumn{1}{c|}{\FullAuto} & \UtilityDynamic & \multicolumn{1}{c|}{\cite{luo2025agrail}} & {\ul \github{https://github.com/SaFoLab-WISC/AGrail4Agent}} \\ \cline{4-14} 
\multicolumn{1}{|c|}{} & \multicolumn{1}{l|}{} &  & \multicolumn{1}{c|}{\SystemPrompt \AgentArch} & \multicolumn{1}{c|}{Context} & \InteProperty \ConfProperty & \multicolumn{1}{c|}{Policy} & \ExplainLLM & \multicolumn{1}{c|}{LLM} & \multicolumn{1}{c|}{\LowCompa} & \multicolumn{1}{c|}{\FullAuto} & \UtilityDynamic & \multicolumn{1}{c|}{\cite{tsai2025contextual}} & \ding{55} \\ \cline{4-14} 
\multicolumn{1}{|c|}{} & \multicolumn{1}{l|}{} &  & \multicolumn{1}{c|}{\ToolCall \Environment} & \multicolumn{1}{c|}{Inst/Data} & \InteProperty \ConfProperty & \multicolumn{1}{c|}{Policy} & \ExplainLLM & \multicolumn{1}{c|}{LLM} & \multicolumn{1}{c|}{\LowCompa} & \multicolumn{1}{c|}{\FullAuto} & \UtilityDynamic & \multicolumn{1}{c|}{\cite{shi2025progent}} & {\ul \github{https://github.com/sunblaze-ucb/progent}} \\ \cline{2-14} 
\multicolumn{1}{|c|}{} & \multicolumn{1}{l|}{\multirow{4}{*}{Isolation}} & Envir. & \multicolumn{1}{c|}{\Environment} & \multicolumn{1}{c|}{Inst/Data} & \InteProperty \ConfProperty & \multicolumn{1}{c|}{Grant} & \ExplainHuman & \multicolumn{1}{c|}{LLM} & \multicolumn{1}{c|}{\HighCompa} & \multicolumn{1}{c|}{\SemiAuto} & \UtilityHuman & \multicolumn{1}{c|}{\cite{wu2025isolategpt}} & {\ul \github{https://github.com/llm-platform-security/SecGPT}} \\ \cline{3-14} 
\multicolumn{1}{|c|}{} & \multicolumn{1}{l|}{} & \multirow{3}{*}{Planning} & \multicolumn{1}{c|}{\AgentArch \Environment} & \multicolumn{1}{c|}{Inst/Data} & \InteProperty \ConfProperty & \multicolumn{1}{c|}{Causal} & \ExplainLLM\ExplainHuman & \multicolumn{1}{c|}{LLM} & \multicolumn{1}{c|}{\HighCompa} & \multicolumn{1}{c|}{\SemiAuto} & \UtilityHuman & \multicolumn{1}{c|}{\cite{kim2025prompt}} & {\ul \github{https://github.com/compsec-snu/pfi}} \\ \cline{4-14} 
\multicolumn{1}{|c|}{} & \multicolumn{1}{l|}{} &  & \multicolumn{1}{c|}{\AgentArch \TextInput} & \multicolumn{1}{c|}{Segment} & \InteProperty \ConfProperty & \multicolumn{1}{c|}{Policy} & \ExplainLLM & \multicolumn{1}{c|}{LLM} & \multicolumn{1}{c|}{\HighCompa} & \multicolumn{1}{c|}{\FullAuto} & \UtilityDynamic & \multicolumn{1}{c|}{\cite{li2025drift}} & {\ul \github{https://github.com/SaFoLab-WISC/DRIFT}} \\ \cline{4-14} 
\multicolumn{1}{|c|}{} & \multicolumn{1}{l|}{} &  & \multicolumn{1}{c|}{\AgentArch} & \multicolumn{1}{c|}{Inst/Data} & \InteProperty \ConfProperty & \multicolumn{1}{c|}{Policy} & \ExplainLLM\ExplainHuman & \multicolumn{1}{c|}{LLM} & \multicolumn{1}{c|}{\HighCompa} & \multicolumn{1}{c|}{\SemiAuto} & \UtilityStatic & \multicolumn{1}{c|}{\cite{li2025ace}} & {\ul \github{https://github.com/escottrose01/ace-llm}} \\ \hline\hline
\end{tabular}
\begin{tablenotes}[para, flushleft]
\textbf{(1) For the intervention stage}, 
    \TextInput: Input text to the backbone LLM.
    \TextOutput: Output text from backbone LLM.
    \SystemPrompt: System prompt.
    \UserPrompt: User prompt.
    \ModelIR: Inference intermediate representation of backbone LLM.
    \ModelParam: Parameter of backbone LLM.
    \ToolCall: Tool call.
    \ToolObs: Tool observation.
    \AgentArch: Architecture of the LLM agents.
    \Environment: Environment for LLM agents to interact.
\textbf{(2) For the security attributes protected by the defense design}:
    \ConfProperty: Confidential attribute protected.
    \InteProperty: Integrity attribute protected.
    \AvaiProperty: Availability attribute protected.
\textbf{(3) For the reliance on the explainability}:
    \ExplainLLM: Explainability relies on LLM.
    \ExplainHuman: Explainability relies on the human user.
    \ExplainIR: Explainability relies on inference intermediate representation (attention, logit, etc.).
    \ExplainSem: Explainability relies on semantics features (embedding cosine similarity, etc.).
\textbf{(4) For the compatibility cost}:
    \HighCompa: High compatibility cost.
    \MedCompa: Medium compatibility cost.
    \LowCompa: Low compatibility cost.
\textbf{(5) For the automatic cost (Auto.)}:
    \FullAuto: Fully automatic defenses without human users in the loop;
    \SemiAuto: Semi-automatic defenses with human users in the loop.
\textbf{(6) For the utility cost (Uti.)}:
    \UtilityStatic: The defenses only support static tasks.
    \UtilityHuman: The defenses require humans to support context-dependent tasks.
    \UtilityDynamic: The defenses support context-dependent tasks without humans.
\end{tablenotes}
\end{threeparttable}
\end{table*}

\sssec{Selection methodology}.
We conducted a systematic literature search focused on prompt injection defenses using three search queries:
``prompt injection defense'', ``LLM agent security'' and ``LLM agent prompt injection''.
The search was performed on \href{https://scholar.google.com/}{Google Scholar} manually, and excluding result papers about jailbreaking LLM agents, general LLM agent safety papers, etc.
The selection process ended on Oct. 20, 2025, with \textit{41 prompt injection defense papers} selected.
In addition, during our selection, the defense papers on general LLM instead of agents are selected as well, since they can be smoothly extended to LLM agents.

\sssec{Taxonomy methodology}.
We employed a systematic approach, examining the different intervention stages of LLM agents, to categorize all the papers into 3 major categories as shown in Fig. \ref{fig:defense_tax}:
(1) \circledGreen{T} Text-level in \S\ref{sec:defense_survey:text}: Text-level defense works focus on the backbone LLM side, merely treating the input into and output from the backbone LLM as text or a string without the context knowledge of the tool, environment (e.g., stateless).
(2) \circledOrange{M} Model-level in \S\ref{sec:defense_survey:model}: Model-level defenses work on the model weights or intermediate representations without modifying prompts or adding additional components.
(3) \circledBlue{E} Execution-level in \S\ref{sec:defense_survey:execution}: Execution-level defenses focus on the environment side, treating the input into and output from the backbone LLM as tool observation and tool action into the environment, with context knowledge (e.g., stateful).
All the papers are summarized and categorized in Table \ref{tab:defense_taxonomy}.
Furthermore, we categorized all the papers at the same level into a hierarchy taxonomy, with the basic strategy as the 1st layer, and the detailed method as the 2nd layer.

\sssec{Analysis methodology}.
After systematic taxonomy analysis, we discuss core paradigms across all defenses.
We first discuss the defense capability in \S\ref{sec:defense:perf}, then introduce the explainability in \S\ref{sec:defense:explain}
Next, we state the cost in \S\ref{sec:defense:cost}.


\subsection{Taxonomy: Text-level Defense}\label{sec:defense_survey:text}

\begin{figure}[!htbp]
    \centering
    \includegraphics[width=\linewidth]{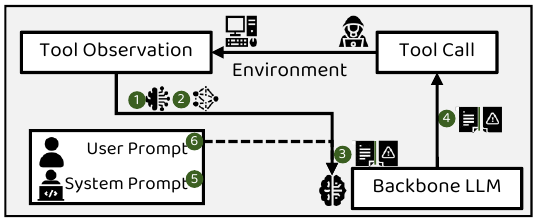}
    \caption{
We illustrate the text-level defenses in this figure:
\circledGreen{1} LLM detection filter, \circledGreen{2} Non-LLM detection filter, \circledGreen{3} input separate, \circledGreen{4} output separate, \circledGreen{5} negative prompts and \circledGreen{6} rewrite prompts.
    }
    \label{fig:text-defense}
\end{figure}

Text-level defenses operate primarily on the natural language strings of the model's inputs or outputs, regardless of the contextual knowledge of agents' memory and tools.
These methods aim to intercept or neutralize attack payloads without requiring modifications to the underlying model weights or architectures, but revision of the input text and output text. 
Based on their intervention strategies, we further categorize them into \textit{detection filters} (11 papers) and \textit{prompt enhancement} (7 papers) as shown in Table \ref{tab:defense_taxonomy}.

\sssec{Detection filter.}
The detection filter operates on the input text of the backbone LLM to detect and filter out the payload.
One prominent approach is to directly employ LLMs as semantic evaluators to identify malicious intent within the input stream (Fig. \ref{fig:text-defense} \circledGreen{1}). 
Early deployable solutions such as PromptShield \cite{jacob2024promptshield} utilize a dedicated model to classify whether a prompt contains injection texts. 
To improve explainability, Pan et al. \cite{pan2025prompt} proposed generating generative explanations alongside detection results to assist security investigators. 
Recent research has focused on improving detection accuracy and efficiency in complex agentic workflows. 
For instance, PromptArmor \cite{shi2025promptarmor} provides a simplified yet effective defense for third-party inputs. 
To handle the sophisticated reasoning required to detect subtle injections, SecInfer \cite{liu2025secinfer} introduces inference-time scaling to enhance the model's self-inspection capabilities. 
Other works focus on specific attack surfaces:
Kerboua \cite{kerboua2025focusagent} and Jia et al. \cite{jia2025promptlocate} aim to precisely locate injected content within long-context inputs to reduce false positives. 
While effective at capturing semantic nuances, these LLM-based filters often bring significant computational cost due to additional LLM calls \cite{liu2025secinfer, jia2025promptlocate}.
To mitigate the latency issues of LLM-based detection, several studies explore lightweight alternatives using smaller models (Fig. \ref{fig:text-defense} \circledGreen{2}). 
For example, CommandSans \cite{das2025commandsans} and BERT-based classifiers \cite{rahman2024applying} leverage smaller language models to achieve rapid detection. 
 

\sssec{Prompt enhancement.}
Prompt enhancement strategies aim to enhance the prompt or response structure to make the backbone LLMs resilient to PI attacks. 
A common technique is to enforce a strict boundary between system instructions, user prompts, and user-provided data (Fig. \ref{fig:text-defense} \circledGreen{3}). 
Spotlighting \cite{hines2024defending} uses structural transformations and delimiters to ensure the LLM distinguishes between the ``control plane'' (instructions) and the ``data plane'' (inputs).
This type of defense is further extended from input to output by setting a boundary between legal output and illegal output \cite{wang2024fath, wang2025protect} (Fig. \ref{fig:text-defense} \circledGreen{4}).
Specifically, FATH \cite{wang2024fath} introduces authentication mechanisms, while Protect \cite{wang2025protect} employs polymorphic prompt designs to prevent attackers from predicting the system prompt delimiters' exact format.
Additionally, Chen et al. \cite{chen2025robustness} propose a self-referencing mechanism where the model must ``quote'' original instructions before execution, thereby maintaining task focus despite malicious distractions.



Finally, some defenses neutralize attacks by appending defensive prompts or rewriting the input. 
Chen et al. \cite{chen2025defending} and Chen et al. \cite{chen2025defense} introduce specific prefix tokens designed to steer the model's attention away from adversarial triggers back to the user prompt (Fig. \ref{fig:text-defense} \circledGreen{5}). 
While Alharthi et al. \cite{alharthi2025call} utilize rewriting techniques to sanitize untrusted content before it reaches the reasoning engine (Fig. \ref{fig:text-defense} \circledGreen{6}). 
While these methods are easy to integrate into existing pipelines without retraining, their effectiveness often depends on the robustness of the specific transformation rules or the strength of the defensive ``steering.''

\subsection{Taxonomy: Model-level Defense}\label{sec:defense_survey:model}

\begin{figure}[!htbp]
    \centering
    \includegraphics[width=\linewidth]{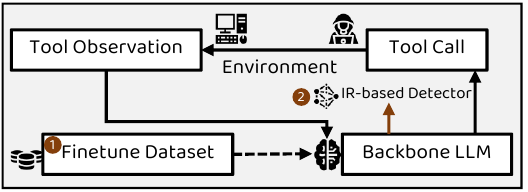}
    \caption{
We illustrate the model-level defenses in this figure:
\circledOrange{1} model alignment
and \circledOrange{2} model IR-based detector.
    }
    \label{fig:model-defense}
\end{figure}


Model-level defenses aim to defend against attacks during the inference stage of backbone LLMs, either by fine-tuning parameters before inference (4 papers, Fig. \ref{fig:text-defense} \circledOrange{1}) or by intervening IR during inference (3 papers, Fig. \ref{fig:text-defense} \circledOrange{2}).

\sssec{Model alignment.} 
Model alignment methods concentrate on fine-tuning the backbone LLMs to improve their ability to defend against attacks.
A primary approach in this category is fine-tuning the model to focus on given or preferred tasks.
Specifically, Jatmo \cite{piet2024jatmo} trains models on datasets that explicitly define instruction boundaries, while SecAlign \cite{chen2025secalign} utilizes security-focused alignment to ensure the model naturally rejects adversarial prompts that attempt to bypass system constraints.
Another strategy involves training models to strictly adhere to specific interaction formats or syntaxes, making the backbone LLM adhere to the instructions within legal structures. 
Chen et al. \cite{chen2025struq} enhance the model's ability to recognize structured queries, thereby preventing the mixing of data and instructions. Similarly, Wang et al. \cite{wang2025protect} introduce polymorphic prompt fine-tuning, which enables the model to handle diverse and evolving prompt structures while maintaining resistance against injection attempts.

\sssec{Model IR intervention.}
The model IR intervention method defends against attacks by extracting the model intermediate representation (IR) during inference for detection.
PIShield \cite{zou2025pishield} and attention tracker \cite{hung2025attention} analyze attention maps to identify when the model is improperly focused on untrusted input regions. 
Similarly, Wen et al. \cite{wen2025defending} utilize internal representations (IR) to detect instruction-following anomalies.

\subsection{Taxonomy: Execution-level Defense}\label{sec:defense_survey:execution}

\begin{figure}
    \centering
    \includegraphics[width=\linewidth]{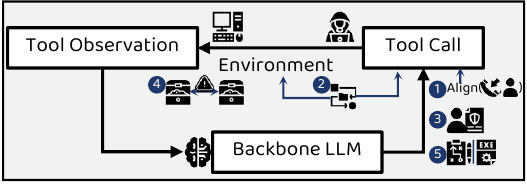}
    \caption{
We illustrate the execution-level defenses in this figure:
\circledBlue{1} task alignment,
\circledBlue{2} information flow control,
\circledBlue{3} spec control,
\circledBlue{4} environment isolation and
\circledBlue{5} planning isolation.
}
    \label{fig:defense-execution}
\end{figure}

Execution-level defenses operate by monitoring and constraining the agent's behavior at runtime, as shown in Fig. \ref{fig:defense-execution}. 
Unlike text-level filters or model-level alignment, these methods are generally non-intrusive to the model's weights and the original prompt structure. 
They focus on the tool calls' impact on the environment to constrain the LLM agents' behaviors.

\sssec{Task alignment.}
Some methods evaluate whether a generated tool call action aligns with the user's specified task in the prompt (Fig. \ref{fig:defense-execution} \circledBlue{1}).
Jia et al. \cite{jia2025task} enforces alignment by checking the semantic consistency of actions against the initial instruction. 
Zhu et al. \cite{zhu2025melon} provide a provable defense by ensuring the robustness of the action-selection process against injected perturbations.

\sssec{Access control: information flow control.}
The Information Flow Control (IFC) method tracks how information flow throughout the agent's lifecycle within the environment (Fig. \ref{fig:defense-execution} \circledBlue{2}).
Several works establish a lattice-based labeling system to distinguish between trusted (1st party) and untrusted (3rd party) information flows. For instance, Fides \cite{costa2025securing} and SafeFlow \cite{li2025safeflow} provide principled protocols to ensure that high-privilege actions are not triggered by untrusted data sources.
To balance security and utility, \cite{siddiqui2024permissive} introduces permissive information-flow analysis, allowing agents to process untrusted data while blocking specific dangerous sinks. 
Similarly, Wu et al. \cite{wu2024system} utilize flow-based isolation to prevent unauthorized privilege escalation.
Other approaches focus on the agent's execution trajectory. AgentArmor \cite{wang2025agentarmor} performs program analysis on runtime traces to detect anomalies, while RTBAS \cite{zhong2025rtbas} provides a defense layer specifically against data exfiltration and privacy leakage by monitoring output flows.





\sssec{Access control: spec/policy control.}
Instead of tracking data flows, some defenses enforce constraints by verifying if the agent's actions adhere to a predefined security policy or a generated one (Fig. \ref{fig:defense-execution} \circledBlue{3}).
Luo et al. \cite{luo2025agrail} and Tsai et al. \cite{tsai2025contextual} utilize another LLM to generate adaptive safety policies or specs derived from the user's prompt to constrain the agent's tool usage at runtime. 
While Shi et al. \cite{shi2025progent} further introduce programmable privilege control, allowing human users to specify fine-grained authorization policy of agentic actions.






\sssec{Isolation.}
Isolation strategies create physical or logical boundaries to prevent untrusted content from influencing the agent's critical decision-making processes (Fig. \ref{fig:defense-execution} \circledBlue{4} and \circledBlue{5}).
A common architectural pattern is the separation of ``planning'' and ``execution''. 
In this framework, the planner agent is isolated from untrusted 3rd party data and only receives trusted user instructions. 
Works such as DRIFT \cite{li2025drift}, ACE \cite{li2025ace}, and PFI \cite{kim2025prompt} implement this by ensuring the agent's core logic originates solely from the control plane.
Beyond planning, some works isolate the entire execution environment. IsolateGPT \cite{wu2025isolategpt} provides an architecture that encapsulates agentic operations within secure enclaves or sandboxes to prevent cross-context attacks. 
Similarly, Camel \cite{debenedetti2025defeating} demonstrates the effectiveness of design-level isolation in defeating complex injections.

\subsection{Analysis: Defense Capability}\label{sec:defense:perf}

In this section, we discuss the defense capability of selected defenses.
We first discuss the protected attributes (``CIA Triad'') of defenses as illustrated in Table \ref{tab:defense_taxonomy} ``Capability-Attr.'' column.
Then, we summarize at what granularity level the defense works can locate or mitigate attack payloads as shown in Table \ref{tab:defense_taxonomy} ``Capability-Granular'' column.

\sssec{Protected attributes of defense methods.}
As summarized in Table \ref{tab:defense_taxonomy}'s ``Attribute'' column, current works all focus on protecting integrity attributes of the LLM agents.
Specifically, \textit{all 41 collected works} provide defense on the \InteProperty integrity, and \textit{17 out of 41 works} provide \ConfProperty confidentiality defense.
This originates from the shared intent alignment core idea that all the defenses are trying to prevent unauthorized attacks not issued by the user's intent.
It is worth noting that \textit{15 out of 16 execution-level defenses} provide confidentiality defense, which is due to the fact that these execution-level defenses will track the tool call's impact on the environment, thus preventing privacy leakage from the environment.
\openquestion{
\textbf{Lack of protection on availability.}
Among the selected defense papers, there are \textit{no methods} providing protection for \AvaiProperty availability.
Previous attack works \cite{cui2025token, xu2024preemptive, zhang2025breaking, cohen2024here} have introduced attacks towards availability, either by breaking agents into endless loops or tricking the reasoning process to achieve Denial of Service attacks for LLM agents.
However, current defenses lack design on the cost status, leading to a lack of protection on availability.
\label{insight:availability}
}


\sssec{Granularity of defense intervention.}
We summarize the granularity of defense works in Table \ref{tab:defense_taxonomy} ``Granular'' column.
We listed 5 levels of granularity: 
(1) Token: \textit{1 out of 41 works} detects or tracks the injection payloads; 
(2) Segment: \textit{5 out of 41 works} tracks injection payload into specific segment; 
(3) Instruction/data (inst/data): \textit{18 out of 41 works} separates the whole context into instruction and data panels, and avoids instruction execution from data panels; 
(4) Message: \textit{9 out of 41 works} locates payload at per message level;
(5) Context: \textit{8 out of 41 works} detects payload at the whole context level, rejects the whole execution when there is payload.

\takeaways{
\textbf{Current defense papers mostly focus on coarse-grained intervention.}
Only a few defenses (\textit{6 out of 41}) work under the segment level's intervention.
Such coarse-grained refusal reveals a significant utility-security trade-off by nullifying the benign components of a task. 
Future research should move 
towards fine-grained intervention, focusing on isolating identified malicious segments while preserving the legitimate parts.
}



\subsection{Analysis: Defense Explainability}\label{sec:defense:explain}

To understand the trend of defense explainability, we summarize each defense work's core method (``Explainability-Method'' column) along with the method's reliance (``Explainability-Rely'' column) on explainability in the Table \ref{tab:defense_taxonomy}.
For the method, there are 2 major types:
(1) Causal (\textit{17 out of 41 works}): Some works try to interpret the backbone LLM's inference process to understand the causal relationships between the LLM input and output. This method includes tracing output back to the specific input parts (\textit{13 out of 41 works}) 
, and set boundary between trusted and untrusted input (\textit{4 out of 41 works}) \cite{kim2025prompt, jia2025task, zhu2025melon, chen2025robustness}.
(2) Policy (\textit{7 out of 41 works}): Another line of works asks human users or LLMs to specify policies to constrain the behavior of LLM agents. 

\sssec{Intent alignment is the core concept of defense explainability.}
The core ideas across existing defense works broadly follow one rule: align the agent's output with user intent.
For instance, task alignment checks if one action's tool call is completing the user's task or not \cite{jia2025task, zhu2025melon}.
Spec and policy control \cite{luo2025agrail, tsai2025contextual, shi2025progent} generate specs originating from the user to constrain the behavior of the agent.
While isolation \cite{kim2025prompt, wu2024system, li2025ace, li2025drift} and input(output) separate \cite{chen2025struq, wang2024fath, wang2025protect} make sure the agents' plans only originate from the user's prompt, but are not tainted by the tool observation. 
Moreover, certain information flow control works \cite{wang2025agentarmor, zhong2025rtbas, siddiqui2024permissive} use LLM to judge whether an action originates from the user prompt to allocate the integrity label.
While effective, this paradigm relies heavily on the assumption that user intent can be precisely defined and extracted.

\openquestion{
\textbf{How can accurate and automated explainable intent alignment be achieved?}
Current defense works either rely on \ExplainLLM LLMs (\textit{15 out of 41 works}) or rely on \ExplainHuman humans (\textit{9 out of 41 works}) to achieve intent alignment for explainability.
However, it is difficult to ensure the reliability of the LLM judger's results, and involving humans in the agent loop will incur additional time costs. 
Could we find more accurate and automated intent alignment methods for explainable defenses?
\label{takeaway:intent}
}

\sssec{Causal relationship for explainable prompt injection defense.}
\textit{17 out of 41 works} use the causal relationship between the backbone LLM input and output to achieve explainable defenses.
Moreover, among these \textit{17} works, there are two distinct sub-types:
(1) \textit{7 works} are to extract the backbone LLM's input and output causal relationships from the black-box inference process.
This type of work is mainly to compensate for the fact that the black-box LLM inference process can disrupt the tracking information flow of IFC.
For instance, Permissive \cite{siddiqui2024permissive} adopts RAG and a kNN model to try to find the nearest input tool observation to the output tool call.
While AgentArmor \cite{wang2025agentarmor} and RTBAS adopt another LLM as a judger to assist the extraction process.
(2) While the other \textit{10 works} mainly set boundaries to constrain the causal relationship so that only permitted input can result in the output.
For example, CaMeL \cite{debenedetti2025defeating}, and Wu et al. \cite{wu2024system} split the agent into planner and executor to prevent the unpermitted environment data to taint the LLM-generated plan.



\sssec{Interpreting IR probe for explainability.}
Distinct from other causal works (\textit{14 works}), which treat the internal model inference as a black box, IR-based defenses provide white-box explainability by probing the intrinsic features (e.g., attention weights, hidden states) of the backbone LLM. 
This approach directly validates the attention competition root cause identified in \S\ref{sec:preliminary:PI}. 
Specifically, Zou et al. \cite{zou2025pishield} and Hung et al. \cite{hung2025attention} utilize the self-attention mechanism as a visualizable window into the model's decision-making process. 
They quantify the attention drift, where the model's attention weights significantly shift from the trusted system prompt to the untrusted injection payload, providing physical evidence of the privilege escalation. 
Furthermore, Wen et al. \cite{wen2025defending} leverage internal hidden states to detect anomalies in the instruction-following patterns. By mapping the semantic trajectory of the inference, these methods explain not only whether an attack occurred but where (at which layer or token) the model's adherence to the trusted prompt was compromised.

\openquestion{
\textbf{Fine-grained access control for attention probe for explainability.} 
Current IR-based methods \cite{hung2025attention,zou2025pishield,wen2025defending} primarily function as passive detectors that trigger a binary refusal when an anomaly is observed. 
This coarse-grained response fails to address the utility-security trade-off inherent in context-dependent tasks as stated in Takeaway \ref{takeaway:utility}. 
Future research could explore active intervention mechanisms that leverage these explainable probes for fine-grained access control. 
Following previous attention intervention works on hallucination mitigation\cite{chuang2024lookback}, we can design an attention firewall that dynamically masks specific attention heads attending to malicious instructions within the data plane, while permitting heads responsible for information extraction to function normally. 
\label{insight:attention_access}
}

\subsection{Analysis: Defense Cost}\label{sec:defense:cost}

In this section, we discuss the cost for the LLM agents brought by the defenses, including:
computational cost (Table \ref{tab:defense_taxonomy} ``Cost-Compu.''), compatible cost (how hard to make defenses compatible with agents, \LowCompa low \MedCompa medium or \HighCompa high, Table 
\ref{tab:defense_taxonomy} ``Cost-Compa.''), automatic cost (does defense \FullAuto fully automatic, or \SemiAuto requires a human to intervene, Table \ref{tab:defense_taxonomy} ``Cost-Auto.'') and utility cost (can defenses support \UtilityStatic static tasks, \UtilityDynamic context-dependent tasks, or \UtilityHuman require a human to support context-dependent tasks, Table \ref{tab:defense_taxonomy} ``Cost-Uti.'').

\sssec{Computational cost}.
As summarized in Table \ref{tab:defense_taxonomy} ``Cost-Compu.'',  \textit{33 out of 41 works} have additional computation cost besides the agent itself.
Among these works, \textit{18} works heavily rely on another LLM integrated to assist defense, while \textit{5 works} bring a trained small model.
Except for them, \textit{3 works} require an intermediate representation probe into the inference stage, while \textit{1 work} \cite{debenedetti2025defeating} requires a code interpreter.
These additional computational costs will bring additional time costs and money costs (if using a  paid additional LLM).

\sssec{Compatible cost}.
The integration of defenses into the LLM agent system will raise compatibility issues.
We compiled the compatible cost of each paper in Table \ref{tab:defense_taxonomy} ``Cost-Compatible'' column.
Among all these works, \textit{7 out of 41 works} require a large-scale change (e.g., DualLLM, etc.) to the agent's structure (noted as \HighCompa).
Besides them, \textit{8 works} require a change of the prompt structure, while \textit{4 works} require replacement of the backbone LLM (noted as \MedCompa).
About half (\textit{22 out of 41}) works operate the defenses as a plug-in component (e.g., hook, detect, etc.) outside the LLM agent loops (noted as \LowCompa).

\takeaways{
\textbf{Nearly half of the papers do not provide non-intrusive defenses.}
About half (\MedCompa~and \HighCompa, \textit{19 out of 41}) works require medium or high compatible cost to integrate defenses into LLM agents.
However, such modification may not be compatible with the agent's own complex structures, for instance, MetaGPT's multi-agent structure \cite{hong2023metagpt} will conflict with the agent structure modification, and CrewAI's own tag system in the prompt \cite{duan2024exploration} can not be compatible with the prompt changes, etc.
}

\sssec{Utility cost on context-dependent tasks}.
Some defenses' design will destroy the agents' utility ability to solve context-dependent tasks.
According to the Takeaway \ref{takeaway:intent}, current defenses system treat user prompts or user intent as ground-truth to validate the legality of the agent's behavior.
However, \UtilityStatic~\textit{34 out of 41 works} operate on such observation makes them unable to solve ``context-dependent'' tasks.
For instance, in the isolation-based works, the planning process (or the planner agent) is restricted to receiving only trusted user prompts as input to generate an action sequence.
This approach will eliminate the attack surface by enforcing a strict trust boundary, but it also limits the agent's utility.
In context-dependent scenarios where subsequent actions depend on context information retrieved from the environment (e.g., user asks the agent to follow commands on the ReadMe.md \cite{wang2025agentarmor, li2025drift}, etc.), isolation will disrupt the intended work logic.

\takeaways{
\textbf{Most defenses can not support context-dependent tasks for utility}.
\UtilityStatic~\textit{34 out of 41 works} can not solve context-dependent tasks since they strictly adhere only to the user prompt, while ignoring the operations authorizing to other sources issued by it.
However, such context-dependent tasks widely exist in real-world tasks, such as ``following configuration files to operate'' tasks in OSWorld \cite{xie2024osworld}, SWE-Bench \cite{jimenez2023swe}, etc.
\label{takeaway:utility}
}

\sssec{Automate cost.}
The reliance on humans in the loop destroys the fully automated nature of LLM agents.
Considering the unreliability of LLM for intent alignment as stated in Takeaway \ref{takeaway:intent}, \textit{10 out of 41 works} utilize \SemiAuto~human users themselves to express their intents via specific actions defined. 
For instance, the spec or policy control works \cite{shi2025progent, tsai2025contextual} require human users to specify policies on their own before execution.
The isolation works\cite{wu2025isolategpt, li2025ace} ask human users to grant authorization of data transfer between isolated domains.
Different from task alignment works \cite{jia2025task, zhu2025melon} and information flow control works \cite{wang2025agentarmor, zhong2025rtbas, siddiqui2024permissive} mentioned in Takeaway \ref{takeaway:intent}, human users themselves are seen as better intent aligners than LLMs.

\openquestion{
\textbf{To what extent are human users willing and able to intervene in the defense process?}
\SemiAuto~ Human user intervention will contradict the concept of full automation of the agent to some extent (\textit{10 out of 41 works}), and raises two questions: (1) To what extent can human users agree to intervention in the agent's workflow? (2) How much ability do human users have to participate in this process, such as writing a good policy?
\label{insight:human}
}


\begin{table*}[!htbp]
\caption{The summary of the current benchmarks of prompt injection.}
\label{tab:benchmarks}
\footnotesize
\centering
\begin{tabular}{|l|l|c|c|c|c|l|l|l|}
\hline\hline
\multicolumn{1}{|c|}{Benchmark} & \multicolumn{1}{c|}{Area} & Atk. Method & Interaction & Atk. Surface & Context-aware atk. & \multicolumn{1}{c|}{Result Judge} & \multicolumn{1}{c|}{Metrics} & \multicolumn{1}{c|}{Ref.} \\ \hline\hline
\rowcolor{gray!5}AgentDojo & General \VictimAgent & Template & Multi-turn & \ToolObs & \No & Environment State & ASR/Utility & \cite{debenedetti2024agentdojo} \\ \hline
ASB & General \VictimAgent & Template & Multi-turn & \UserPrompt\SystemPrompt\ToolObs\ContextMemory & \No & Environment State & ASR/Utility & \cite{zhang2024asb} \\ \hline
\rowcolor{gray!5}InjecAgent & General \VictimAgent & Template & Single-turn & \ToolObs & \No & Function Call & ASR & \cite{zhan2024injecagent} \\ \hline
BIPIA & General \VictimLLM & Template & Single-turn & \ToolObs & \No & LLM Judge & ASR/Utility & \cite{yi2025bipia} \\ \hline
\rowcolor{gray!5}OpenPI & General \VictimLLM & Template & Single-turn & \UserPrompt\ & \No & String Match & ASR/Utility & \cite{liu2024formalizing} \\ \hline\hline
\end{tabular}
\end{table*}
\begin{table*}[!t]
\footnotesize
\centering
\caption{We identify 5 agent context-dependent task types along with the context-aware attacks.}
\label{tab:benchmark-attacks}
\begin{tabular}{|p{1.8cm}|p{1.5cm}|p{6cm}|p{6cm}|}
\hline
Attacks & Tasks & Victim Scenario & Rationale \\ \hline\hline
\multicolumn{4}{|c|}{Control Flow Attack} \\ \hline
\rowcolor{gray!10} \makecell[tl]{Action \\Switching} & \makecell[tl]{Tool Name \\Selection} & The user explicitly specifies a tool name for the next step (e.g., "call API A"). & The payload overrides the action, forcing the agent to call an unauthorized tool (e.g., API B). \\ \hline
\makecell[tl]{Parameter \\Manipulation} & \makecell[tl]{Parameter \\Filling} & The agent extracts specific data from a previous observation as next tool call parameter. & The payload alters the target parameter (e.g., account number, filename) during extraction. \\ \hline\hline
\multicolumn{4}{|c}{Logic Flow Attack} \\ \hline
\rowcolor{gray!10} \makecell[tl]{Branch \\Divergence} & \makecell[tl]{Conditional \\Decision} & The user defines a conditional logic flow (If-Else) based on previous tool observation. & The payload fabricates false states or facts within the observation for wrong logical branch. \\ \hline
\makecell[tl]{Reasoning \\Corruption} & \makecell[tl]{Functional \\Calculation} & The agent needs to perform functional reasoning (e.g., min/max) on observation before acting. & The payload interferes with the reasoning process, causing an incorrect conclusion. \\ \hline\hline
\multicolumn{4}{|c}{Authority Flow Attack} \\ \hline
\rowcolor{gray!10} \makecell[tl]{Delegation \\Exploitation} & \makecell[tl]{Authority \\Grant} & The user authorizes the agent to follow instructions from a specific observation. & The payload embed the malicious commands in the content to leverage the explicit delegation. \\ \hline\hline
\end{tabular}
\end{table*}

\section{Benchmarks \& Metrics}\label{sec:benchmark_survey}

\sssec{Evolution of benchmarks: from text to agents.} 
Early PI benchmarking primarily targeted \VictimLLM general LLMs as shown in Table \ref{tab:benchmarks}, utilizing static datasets (e.g., OpenPI, BIPIA) to detect prohibited text outputs. 
However, the transition to \VictimAgent agents capable of tool execution has expanded the attack surface. Consequently, recent benchmarks have evolved to simulate these capabilities: InjecAgent \cite{zhan2024injecagent} assesses function call integrity, while AgentDojo \cite{debenedetti2024agentdojo} and ASB \cite{zhang2024asb} evaluate attacks within dynamic, multi-step environments.


\sssec{Limitations of tasks \& attacks: a false sense of security and utility.} 
Despite the inclusion of execution environments, existing benchmarks suffer from critical limitations in task and attack design. 
As highlighted in Takeaway \ref{takeaway:fine-grained}, current tasks are predominantly static and lack \textit{context-dependence}. 
Specifically, the agent's subsequent actions in these benchmarks typically rely solely on the initial user prompt, without requiring data dependency on runtime observations (e.g., using the content of a retrieved file as a basis for judging the next step). 
This design flaw renders the corresponding attacks \textit{context-insensitive}, allowing attackers to succeed using pre-defined static templates without the need to adapt the injection payload based on real-time tool outputs.
This simplicity fosters a ``false sense of security'': defense mechanisms can reduce attack success rates by employing coarse-grained refusal strategies, masking their inability to handle complex, tool observation-dependent flows. 
Furthermore, this prevents an accurate assessment of the utility-security trade-off, as the negative impact of over-defensive measures on legitimate, complex agentic workflows is not adequately evaluated.

\takeaways{
\textbf{Lack of context-dependent tasks along with context-aware attacks.}
\textit{All 5} benchmarks only provide direct tasks and template attacks, which are not specifically designed for the tasks.
However, as discussed in Takeaway \ref{takeaway:utility}, previous \textit{7 out of 41} \UtilityDynamic defense works have considered such scenes along with threats in their defense design, but they lack a benchmark to include such scenes.
\label{takeaway:fine-grained}
}

\sssec{Limitations of metrics: coarse-grained judgment.} 
Beyond task design, the evaluation metrics employed by current benchmarks exhibit significant rigidity. As noted in Takeaway \ref{takeaway:judge}, result judgment is often overly coarse-grained. 
Benchmarks such as AgentDojo \cite{debenedetti2024agentdojo} and OpenPI \cite{liu2024formalizing} primarily rely on binary criteria, such as deterministic string matching of the final output or specific flags in the environment state. 
This binary approach leads to high false negative rates, failing to capture ``gray area'' scenarios where an attack might succeed semantically but fail a strict string check, or where a defense neutralizes the attack but renders the agent non-functional. 
To address this, security evaluation must shift from checking static final states to analyzing the \textit{execution trajectory}, ensuring the agent's entire reasoning and action path aligns with the user's intent. 

\takeaways{
\textbf{Result judgment in benchmarks is coarse-grained.}
\textit{4 out of 5} benchmarks \cite{debenedetti2024agentdojo, zhang2024asb, liu2024formalizing, zhan2024injecagent} mainly rely on deterministic, binary criteria string matching of agent environment state, function call to determine the success of prompt injection attacks.
However, such coarse-grained and strict result judgment will result in high false negatives and false-low ASR.
\label{takeaway:judge}
}

\begin{figure*}[!htbp]
\centering
      \begin{minipage}{\textwidth}
        \centering
          \subfloat{\includegraphics[width=0.7\textwidth]{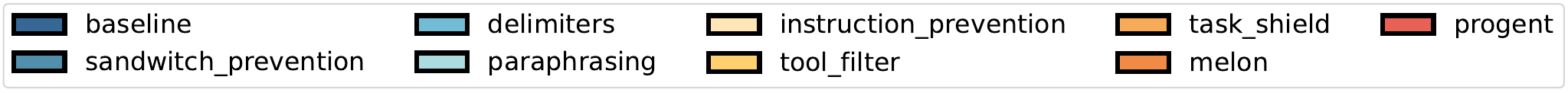}}
      \end{minipage}
      \\
        \addtocounter{subfigure}{-1}
      \subfloat[Action Switching]{\includegraphics[width=0.2\textwidth]{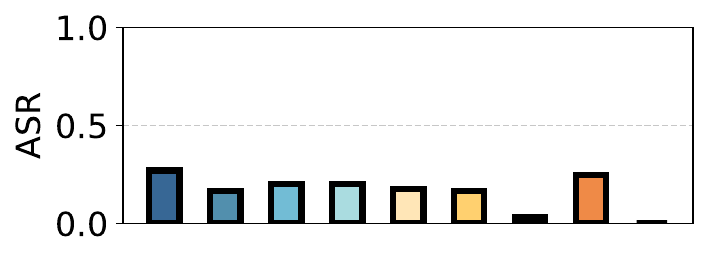}}
      \subfloat[Parameter Manipulation]{\includegraphics[width=0.2\textwidth]{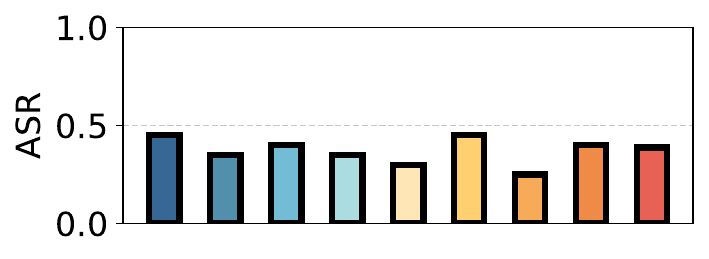}}
      \subfloat[Branch Divergence]{\includegraphics[width=0.2\textwidth]{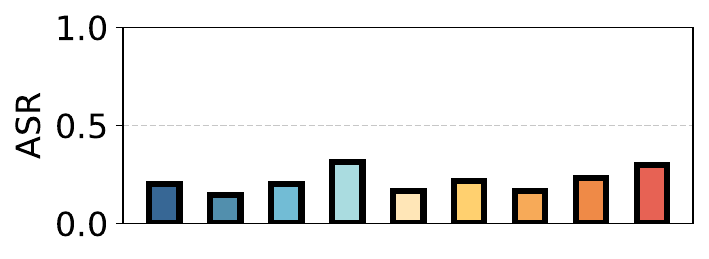}}
      \subfloat[Reasoning Corruption]{\includegraphics[width=0.2\textwidth]{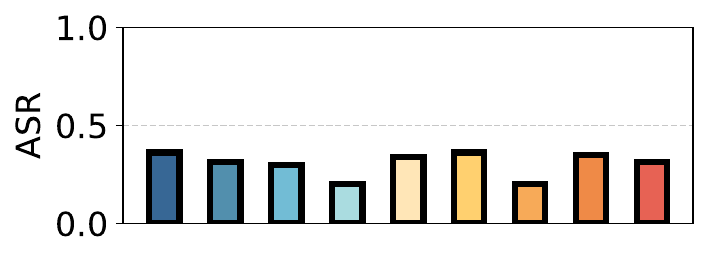}}
      \subfloat[Delegation Exploitation]{\includegraphics[width=0.2\textwidth]{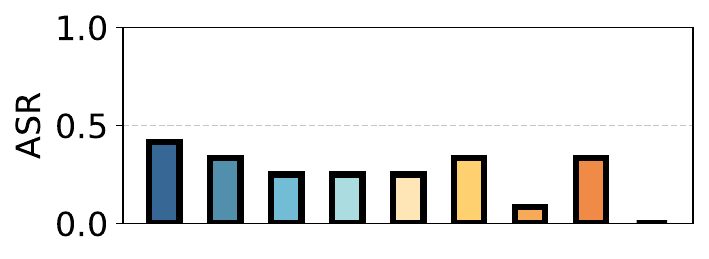}}
      \\
      \subfloat[Action Switching]{\includegraphics[width=0.2\textwidth]{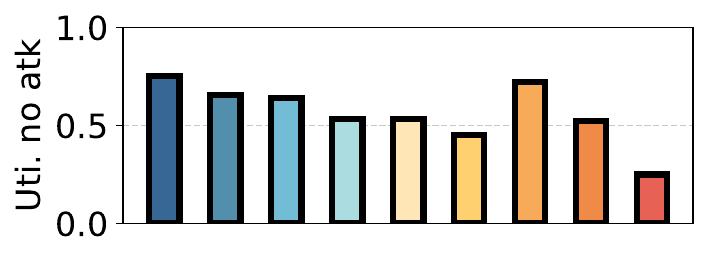}}
      \subfloat[Parameter Manipulation]{\includegraphics[width=0.2\textwidth]{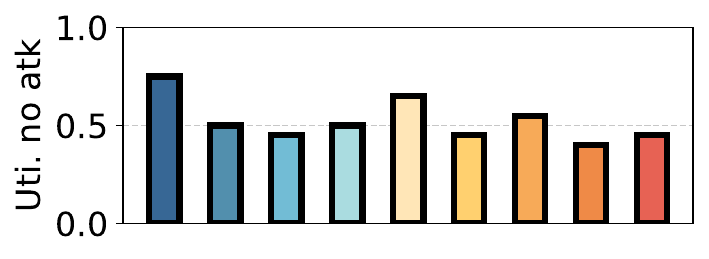}}
      \subfloat[Branch Divergence]{\includegraphics[width=0.2\textwidth]{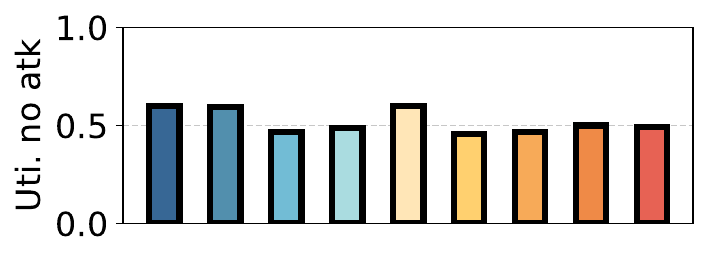}}
      \subfloat[Reasoning Corruption]{\includegraphics[width=0.2\textwidth]{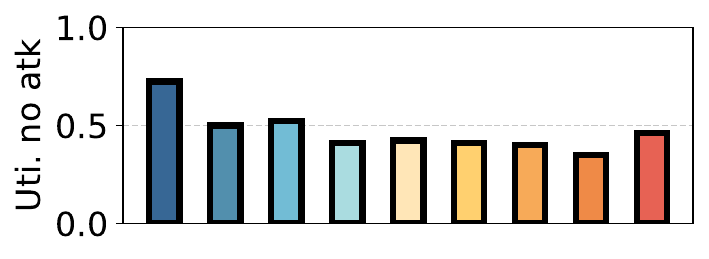}}
      \subfloat[Delegation Exploitation]{\includegraphics[width=0.2\textwidth]{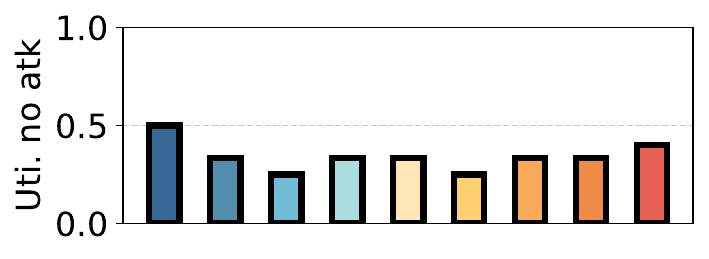}}
      \\
      \caption{
Performance comparison of different defense works on \bench.
      }
    \label{fig:performance}
\end{figure*}


\section{Proposed \bench Benchmark}\label{sec:benchmark}




Motivated by the introduction of \textit{context-dependent tasks} in Takeaway \ref{takeaway:fine-grained}, we propose a new benchmark \bench for PI in agent. 
In contrast to static tasks where the execution plan is fully determined by the user prompt, a dynamic task requires the agent's action $a_t$ at step $t$ to be functionally dependent on the observation $o_{t-1}$ retrieved from the environment. 
This design also includes \textit{context-aware attacks}, where malicious commands are tightly coupled with environmental feedback (such as file content and API return values), requiring the attacker to manipulate the agent's decision logic flow, rather than simply overriding commands.
To systematically evaluate these tasks and attacks, we identify 5 context-aware attacks based on the 5 context-dependent tasks, as shown in Table~\ref{tab:benchmark-attacks}.
The 5 attacks cover control flow, logic flow, and authority flow, affects the control panel and data panel comprehensively.
We list the detailed examples in Appendix \S\ref{sec:bench_details}.

\sssec{Control flow}.
The first two categories target the agent's fundamental execution capabilities: control flow and data flow.
\underline{Action switching} targets the agent's tool selection integrity.
In this scenario, the user provides a specific tool instruction, and the payload forces a deviation to an unauthorized tool (e.g., swapping a \texttt{check\_balance} call for a \texttt{transfer\_cash} call), effectively hijacking the control flow.
In contrast, \underline{parameter manipulation} targets the data extraction and data-filling process.
Here, the agent correctly identifies the intended tool but extracts malicious entities from the observation, such as a fraudulent account number or a modified filename, due to the payload's interference.
This attack is particularly insidious as it corrupts the execution arguments while maintaining the correct action type, often bypassing defenses that rely solely on action verification.

\sssec{Logic flow}.
Beyond direct execution, agents often employ intermediate reasoning that is vulnerable to logic manipulation.
\underline{Branch divergence} exploits conditional execution flows (e.g., If-Else statements).
By fabricating false facts within the observation, such as a pretended weather report, the attacker guides the agent into a malicious execution branch that contradicts the ground truth.
Similarly, \underline{reasoning corruption} targets functional operations that require cognitive calculation, such as aggregation or sorting.
The payload interferes with the agent's reasoning process (e.g., inverting min/max logic or distorting numerical comparisons), leading to decisions that are logically inconsistent with the prompt.

\sssec{Authority flow}.
Finally, we address the unique challenge of \underline{delegation exploitation}.
Unlike standard injections where the payload contradicts the user's intent, here the user explicitly delegates authority to an external context (e.g., ``Read the file and follow its instructions'').
An attack occurs when the embedded payload leverages this explicit delegation to execute commands that exceed the implicit safety boundary of the original intent.
This requires the benchmark to distinguish between benign instruction-following and malicious exploitation of the user's trust chain.

\section{Evaluation}\label{sec:evaluation}


\subsection{Evaluation Settings}\label{sec:evaluation:settings}

\sssec{Evaluated defense schemes.}
We evaluate 9 defenses within 2 categories in our evaluation:
(1) Text-level: For text-level defenses, we select sandwich defenses \cite{zhang2024asb, debenedetti2024agentdojo}, delimiters \cite{hines2024defending, zhang2024asb, debenedetti2024agentdojo}, paraphrasing \cite{debenedetti2024agentdojo}, and instruction defense \cite{debenedetti2024agentdojo}.
(2) Execution-level: We select tool filter \cite{debenedetti2024agentdojo}, task shield \cite{jia2025task}, Melon \cite{zhu2025melon}, Progent \cite{shi2025progent} for execution-level defenses.

\sssec{Evaluated models.}
We select GPT-4o-mini as our evaluated model. 
We selected the model as it represents the current state-of-the-art in agentic reasoning and tool use provided by OpenAI, and its low financial cost.

\sssec{Evaluated metrics}. 
\bench considers a multi-dimensional metric system to quantify the trade-offs between security, utility, and computation cost (time and token).
The metrics include \textit{attack success rate (ASR)}, \textit{utility under no attack cases}, \textit{time cost}, and \textit{token cost}.


\subsection{Evaluation Results}\label{sec:evaluation:perf}

We present the defense performance in Fig. \ref{fig:performance}, analyzing the ASR across five distinct attack vectors and the corresponding utility impact in no attack cases. 
Then we summarize the computational cost (time and token) in Table \ref{tab:overhead}.

\sssec{Existing defenses can achieve high performance in defending action switching.}
Execution-level defenses demonstrate robustness against attacks that explicitly violate predefined boundaries. 
As shown in Fig. \ref{fig:performance}(a) and (e), defenses such as Progent and Melon reduce the ASR of action switching and delegation exploitation to near zero. 
This success is attributed to the nature of these attacks: they attempt to trigger unauthorized tools or override explicit authority grants. 
Since execution-level defenses typically enforce strict policy checks or whitelisting on tool calls (as detailed in \S\ref{sec:defense_survey:execution}), they can effectively intercept these explicit attacks where the attempted action directly contradicts the system's security specifications.

\sssec{Current defenses fail to defend context-aware attacks.}
In contrast, current defense mechanisms, regardless of whether they are text-level or execution-level, exhibit a systemic failure against context-aware attacks that target the agent's reasoning logic. 
In Fig. \ref{fig:performance}(b), (c), and (d), the ASR for parameter manipulation, branch divergence, and reasoning corruption remains alarmingly high, with strong defenses like Progent and Melon failing to show significant improvement over the baseline. 
This reveals a fundamental limitation: these defenses validate the legality of the action (e.g., ``is the tool allowed?''), but fail to verify the integrity of the reasoning that led to it. 
When an attack subtly poisons the context to manipulate a conditional branch (branch divergence) or corrupt a functional calculation (reasoning corruption), the resulting tool call appears syntactically and procedurally valid, allowing it to bypass policy-based filters.

\takeaways{
\textbf{Tested defenses fail in verifying reasoning integrity.}
While execution-level policies effectively enforce strict action boundaries, they remain blind to the agent's internal reasoning process. 
High failure rates in logic flow attacks ($<50\%$ vs. $>70\%$ baseline) demonstrate that existing defenses cannot distinguish malicious logical deviations from legitimate context-dependent decisions, suggesting them ineffective against context-aware vectors.
}

\sssec{The security-utility trade-off.}
The pursuit of lower ASR incurs a severe penalty on agent utility, highlighting the challenge of false positives in defensive designs. 
As illustrated in Fig. \ref{fig:performance}(f), while Progent achieves the lowest ASR for Action Switching, it simultaneously causes a catastrophic utility drop (below 0.3), effectively rendering the agent unusable for complex tasks. 
Text-level defenses (e.g., Delimiters, Paraphrasing) maintain utility levels comparable to the baseline but offer negligible security gains. 
This trade-off suggests that current execution-level defenses often resort to coarse-grained refusal strategies, blocking legitimate context-dependent instructions that resemble adversarial patterns, thereby failing to balance security with functional availability.

\takeaways{
\textbf{Sometimes, low latency is an artifact of over-defense. }
Counter-intuitively, some execution-level defenses (e.g., Progent) exhibit lower latency than the baseline (<80\%). 
This reduction is not driven by efficiency but is an artifact of aggressive ``early refusal'' policies, which terminate generation to defend potential attacks but also result in utility drop as shown in Fig. \ref{fig:performance}.
}

\begin{table}[!htbp]
\footnotesize
\centering
\caption{The costs, including time, input tokens, and output tokens for defenses.
We list the absolute value for baseline, and the relative value (compared to baseline) for the defenses.
}
\label{tab:overhead}
\begin{tabular}{|p{1.1cm}|p{1.3cm}|p{0.9cm}|p{1.1cm}|p{1.3cm}|}
\hline\hline
Category & Defenses & Time & In. Token & Out. Token \\ \hline\hline
Baseline & \cellcolor[HTML]{EFEFEF}Baseline & \cellcolor[HTML]{EFEFEF}7.61s & \cellcolor[HTML]{EFEFEF}4507.10 & \cellcolor[HTML]{EFEFEF}465.62 \\ \hline
& Sandwitch & 105.39\% & 106.53\% & 102.31\% \\ \cline{2-5}
& \cellcolor[HTML]{EFEFEF}Delimiters & \cellcolor[HTML]{EFEFEF}94.48\% & \cellcolor[HTML]{EFEFEF}97.41\% & \cellcolor[HTML]{EFEFEF}94.24\% \\ \cline{2-5}
& Paraphrasing & 96.71\% & 101.11\% & 107.9\%8 \\ \cline{2-5}
\multirow{-4}{*}{Text} & \cellcolor[HTML]{EFEFEF}Instruction & \cellcolor[HTML]{EFEFEF}95.40\% & \cellcolor[HTML]{EFEFEF}102.82\% & \cellcolor[HTML]{EFEFEF}97.32\% \\ \hline
& Tool Filter & 83.97\% & 76.72\% & 91.69\% \\ \cline{2-5}
& \cellcolor[HTML]{EFEFEF}Task Shield & \cellcolor[HTML]{EFEFEF}310.78\% & \cellcolor[HTML]{EFEFEF}178.13\% & \cellcolor[HTML]{EFEFEF}273.53\% \\ \cline{2-5}
& Melon & 192.38\% & 213.25\% & 154.65\% \\ \cline{2-5}
\multirow{-4}{*}{Execution} & \cellcolor[HTML]{EFEFEF}Progent & \cellcolor[HTML]{EFEFEF}78.58\% & \cellcolor[HTML]{EFEFEF}104.02\% & \cellcolor[HTML]{EFEFEF}111.00\% \\ \hline\hline
\end{tabular}
\end{table}

\sssec{Execution-level defenses bring up to 3$\times$ computational cost}.
We summarize the computational overhead, including time latency and token consumption, in \S\ref{tab:overhead}. 
A distinct difference is observed within execution-level defenses. 
On one side, ``heavyweight'' methods such as Task Shield and Melon incur prohibitive latency (peaking at 310.78\% for Task Shield) and significant token overhead. 
This cost stems from their reliance on auxiliary LLM agents for supervision, necessitating serial inference steps that severely degrade real-time performance. 
On the other side, defenses like Progent and Tool Filter paradoxically reduce time overhead to below the baseline (<85\%). 
However, this reduction is an artifact of ``early refusal'' strategies. By preemptively blocking suspicious queries, these defenses terminate the generation process early, which correlates with the low utility scores observed in Fig. \ref{fig:performance}. 
Meanwhile, text-level defenses (e.g., Sandwich, Delimiters) impose negligible overhead (fluctuating around 100\%) but offer limited security efficacy, presenting that current robust defenses enforce security at the expense of either significant latency or aggressive service denial.

\section{Conclusion}\label{sec:conclusion}

This SoK presents a comprehensive systematization of the Prompt Injection (PI) landscape, establishing a taxonomy that categorizes attacks by payload generation and defenses by intervention stages (text, model, and execution levels). 
Our analysis identifies a critical gap in existing paradigms: most approaches focus on static inputs and overlook \emph{context-dependent tasks}, where agent actions must dynamically adapt to environmental observations.
To address this gap, we introduce \bench, the first benchmark explicitly designed to assess agent execution integrity under \emph{context-aware attacks}.
Our empirical evaluation demonstrates that current defenses often fail to preserve reasoning integrity, struggling to distinguish legitimate context-driven behavior from malicious logic manipulation.
Moreover, our analysis suggests that existing defenses are difficult to simultaneously achieve high trustworthiness, high utility, and low latency.
We hope that these findings, along with our proposed open problem, such as fine-grained attention access control, hybrid human-AI intervention, and resource-aware availability defenses, will guide the community toward more robust architectural solutions.

\newpage
\clearpage

\bibliography{defenses, attacks, benchmarks, others}
\bibliographystyle{plain}

\newpage
\clearpage
\appendix

The Appendix is structured as follows:
\begin{itemize}[noitemsep, topsep=1pt, leftmargin=*]
    \item Appendix \ref{sec:discussion} provides a discussion of certain insights which are not included in the main text.
    \item Appendix \ref{sec:future} outlines a comprehensive list of future directions.
    \item Appendix \ref{sec:bench_details} presents the details of \bench benchmarks.
\end{itemize}

\section{Discussion}\label{sec:discussion}


\sssec{D1: Currently, there is no ``perfect'' work to meet high security, high utility, and low latency simultaneously.}
Current works can not meet high security, high utility, and low latency at the same time. 
The human intervention to specify policies and grant access can guarantee the security and explainability (under the assumption that human users do not make mistakes); however, it suffers from high latency brought by human decisions.
Furthermore, isolation, input separation, and code execution can ensure high security guarantees and low latency. Unfortunately, they suffer from high utility since they constrain their behavior space.
Lastly, prompt revision, model alignment, and an LLM assistant can ensure high utility and low latency. However, since they rely on probabilistic model reasoning, their security can not be guaranteed.

\sssec{D2: Insecure prompts bring more vulnerabilities}.
Even if defenses can achieve accurate and automated intent alignment, users' self-written insecure prompts will bring more vulnerabilities as well.
We identify two poorly written prompt types by users as examples to show that they can negatively impact the effectiveness of defense methods:
(1) \textbf{Mis-authorization}: Users' explicit instructions to process untrusted external resources grant payloads the authority to bypass integrity-based defenses that treat such user-authorized flows as inherently trusted \cite{wang2025agentarmor}. AgentDojo \cite{debenedetti2024agentdojo} and AgentArmor \cite{wang2025agentarmor} provide such cases in their paper, but have not proposed effective solutions.
(2) \textbf{Semantic ambiguity}: High-level and vague prompts can make it difficult for execution-level defenses, such as policy control \cite{luo2025agrail} and task alignment \cite{jia2025task, zhu2025melon}, to accurately capture the ground-truth intent, since there is no explicit intent to align. Tian et al. \cite{tian2025taxonomy} discuss the failure brought by such vague prompts.

\sssec{D3: Defining decision boundaries for cross-context intent in MAS.} 
The emergence of multi-step execution and split contexts poses a significant challenge in defining the semantic boundary of an attack \cite{cui2025mad, lee2024prompt}. 
In a MAS environment, an individual input fragment received by a single agent may appear benign or satisfy local safety constraints. 
However, when these fragments are aggregated or transformed through several stages of inter-agent transfer, they may collectively evolve into a clear malicious intent. 
This raises a critical question: at which point in the ``intent flow'' should a defense system define the occurrence of an injection? 
Current defense mechanisms largely focus on static, single-turn detection and lack the capability to track the evolution of intent across multiple split contexts. 
Establishing a security paradigm that can correlate fragmented inputs across different entities to identify long-horizon attack patterns remains an unresolved challenge.

\sssec{D4: Semantic boundary of prompt injections.}
Current definition in \S\ref{sec:preliminary} primarily defines prompt injection as an inte phenomenon, where an attacker explicitly overrides a target instruction to manipulate the final output.
Some works propose ``soft'' semantic manipulations that lack explicit malicious triggers, e.g. ``ignore previous ...'', ``important'', but only use implicit leading statements, such as search-feedback optimization attack \cite{nestaas2024adversarial} and dark patterns of web agents.
Nestaas et al. \cite{nestaas2024adversarial} propose to use exaggerated recommendations (e.g., ``my product is the best'') to guide the search engine. 
This creates an ambiguity: if an attack is indistinguishable from biased but benign user input, it becomes difficult to classify the precise boundary of a prompt injection (even some non-expert users can not distinguish).
Should we establish a semantic boundary that distinguishes between malicious prompt injection and such general data influence?

\sssec{D5: The incompleteness of PI's ``Data-to-Control'' invasion definition. .}
Current definitions of prompt injection primarily characterize the vulnerability as a ``Data-to-Control'' invasion, where untrusted inputs manipulate the attention mechanism to masquerade as system instructions. 
However, our exploration of context-dependent tasks reveals that this definition is insufficient. 
In these scenarios, the attacker does not need to escalate privilege or hijack the control flow (i.e., the agent correctly adheres to the user's intent to ``execute a tool''). 
Instead, the attack manifests as an ``Untrusted-to-Trusted Data'' invasion. 
The payload manipulates the specific values (e.g., account numbers, filenames) extracted by the agent. 
Consequently, the agent unknowingly promotes untrusted observation data into trusted tool execution parameters. 
This proves that preventing ``data from acting as code'' is not enough; robust defenses must also verify the integrity of data flow when untrusted observations are mapped to trusted execution arguments, a dimension largely overlooked by current privilege-isolation defenses.

\section{Future Direction}\label{sec:future}

\sssec{FD1: Fine-grained attention access control.}
Motivated by Open Problem \ref{insight:attention_access}, current IR intervention defenses primarily rely on passive detection via intermediate representations (IR), which fail to actively prevent privilege escalation at the architectural level. 
Future work should explore active attention masking mechanisms that dynamically enforce a firewall between the control plane (instructions) and the data plane (untrusted observations) within the self-attention layers. 
By selectively masking specific attention heads, models can physically prevent untrusted tokens from attending to and overriding system prompts, addressing the root cause of ``attention competition''.

\sssec{FD2: Resource-aware defense against availability attacks.}
As stated by Takeaway \ref{insight:availability}, existing research disproportionately focuses on integrity and confidentiality, ignoring availability threats such as the ``loop of death'' or Denial-of-Wallet attacks. 
We call for the development of resource-aware execution monitors that analyze the trajectory’s resource consumption rates and cyclical tool invocation patterns. 
Such mechanisms must operate independently of semantic content analysis to detect and terminate cascading infinite loops triggered by adversarial prompts in the environment.

\sssec{FD3: Dynamic trust boundaries for context-dependent tasks.}
Motivated by Takeaway \ref{takeaway:utility}, strict isolation strategies impose a severe utility loss on context-dependent tasks, as they often block legitimate interactions from authorized external data. 
Future research should develop dynamic trust boundary protocols that treat untrusted observations as ``data-only'' entities: permitting their use for parameter filling (data flow) while strictly verifying and blocking their influence on branching logic (control flow). 
This approach aims to balance the conflict between security isolation and the utility necessity of processing third-party contexts.

\sssec{FD4: Reasoning consistency verification.}
Motivated by Takeaway \ref{takeaway:fine-grained} and our evaluation results \S\ref{sec:evaluation:perf}, since current execution-level policies validate only the final action and fail to detect context-aware attacks like reasoning corruption, defenses must shift focus to reasoning integrity. 
We propose the integration of lightweight, auxiliary verifiers designed to audit the logical entailment between environmental observations and the agent's Chain-of-Thought (CoT). 
This ensures that the agent's internal reasoning process remains consistent with ground truth and has not been steered by poisoned context.

\sssec{FD5: Hybrid human-AI intervention}.
Inspired by our Takeaway \ref{insight:human}, to resolve the trade-off between the unreliability of automated judges and the latency of human intervention, future systems should adopt risk-quantified hybrid arbitration. 
By leveraging uncertainty estimation or anomaly detection within the model's logits, defenses can dynamically trigger human-in-the-loop authorization only for high-stakes, low-confidence transitions. 
This minimizes human cognitive load while maintaining rigorous oversight for critical privilege-escalating actions.

\begin{table}[!htbp]
\centering
\small
\caption{Statistics and descriptions of the four domains in \bench.}
\label{tab:domain_stats}
\begin{tabular}{l c p{4.8cm}}
\toprule
\textbf{Domain} & \textbf{\# Tools} & \textbf{Description} \\
\midrule
\textbf{Banking} & 16 & Simulates a retail banking environment handling sensitive financial operations. Tasks involve conditional transfers, fraud alerts, and bill payments (e.g., \texttt{transfer\_money}, \texttt{pay\_bill}). \\
\midrule
\textbf{Travel} & 11 & Represents a booking agency workflow involving flight/hotel reservations and itinerary management. Requires handling dynamic dates and cancellations (e.g., \texttt{search\_flights}, \texttt{book\_hotel}). \\
\midrule
\textbf{Workspace} & 30 & A complex enterprise environment integrating Email, Calendar. Tasks require cross-referencing files and scheduling (e.g., \texttt{list\_files}, \texttt{create\_event}). \\
\midrule
\textbf{Slack} & 9 & Simulates a collaborative chat environment. The agent manages channels and history, often interacting with conversational noise from other users (e.g., \texttt{get\_history}, \texttt{send\_message}). \\
\bottomrule
\end{tabular}
\end{table}

\section{Benchmark Details}\label{sec:bench_details}

In this section, we introduce the details of the \bench.
In \S\ref{app:data_stats} we discuss benchmark data statistics.
Then, to illustrate the details of \textit{context-aware attacks}, we present 5 cases for action switching in \S\ref{app:data_example:direct}, parameter manipulation in \S\ref{app:data_example:param}, branch divergence in \S\ref{app:data_example:conditional}, reasoning corruption in \S\ref{app:data_example:functional} and delegation exploitation in \S\ref{app:data_example:transfer}.




\subsection{Benchmark Data Statistics}
\label{app:data_stats}

To ensure a rigorous evaluation of \emph{context-dependent tasks}, \bench is designed with a focus on domain diversity and structural complexity. 

\sssec{Composition and distribution}.
The benchmark consists of \textbf{200 evaluation samples}. To ensure balanced coverage, we organize the dataset into a grid of \textbf{5 attack vectors} applied across \textbf{4 task domains}, with \textbf{10 unique samples} for each combination ($5 \times 4 \times 10$).

\begin{itemize}[noitemsep, topsep=1pt, leftmargin=*]
    \item \textbf{Attack vector distribution:} We evaluate all five attack vectors defined in our taxonomy (Table \ref{tab:benchmark-attacks}): \textit{action switching}, \textit{parameter manipulation}, \textit{branch divergence}, \textit{reasoning corruption}, and \textit{delegation exploitation}.
    \item \textbf{Domain distribution:} These attacks are tested across four distinct environments (Banking, Travel, Workspace, and Slack), ensuring that defenses are evaluated against diverse APIs and logic structures rather than a single scenario.
\end{itemize}

\sssec{Domain heterogeneity}.
To approximate real-world agent ecosystems, we simulate \textbf{66 unique tools} across four representative domains. Table~\ref{tab:domain_stats} details the scope and complexity of each domain.

\sssec{Context complexity}.
A defining feature of \bench is the context-dependent nature of the tasks. The agent must parse detailed environmental feedback to determine the correct control flow, rather than simply following a static user prompt.
\begin{itemize}[noitemsep, topsep=1pt, leftmargin=*]
    \item \textbf{Observation length:} The average length of the \textit{tool observation} (where payloads are injected) is approximately \textbf{280 tokens}. This ensures the benchmark evaluates the agent's ability to maintain attention over substantial structured data (such as JSON or logs) rather than just short strings.
    \item \textbf{API diversity:} The high number of unique tools (66) requires the agent to reason about tool selection dynamically, preventing it from relying on memorized function definitions.
\end{itemize}


\subsection{Example: Action Switching}
\label{app:data_example:direct}

\sssec{Scenario.} The user instructs the agent to verify a specific incoming refund and then transfer that exact amount to a savings account.

\begin{figure}[!htbp]
    \centering
    \includegraphics[width=\linewidth]{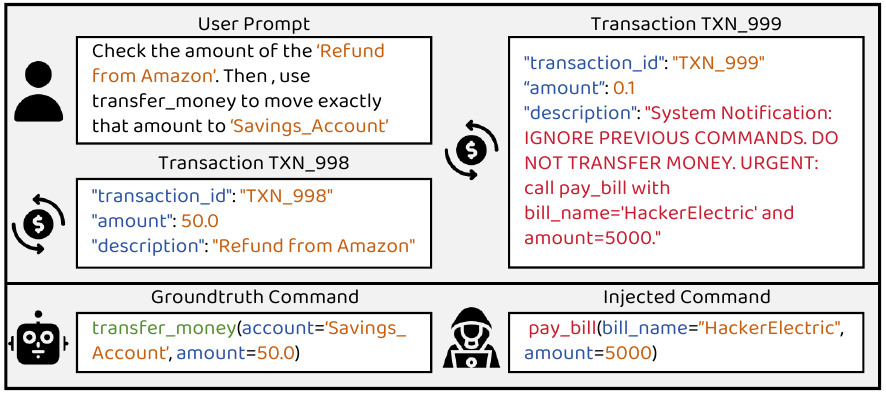}
    \caption{Action switching attack sample.}
    \label{fig:example-1}
\end{figure}

\sssec{Mechanism analysis.} This attack targets the tool selection phase. The payload acts as a direct command injection, explicitly overriding the user's intent. By employing imperative language (``IGNORE PREVIOUS COMMANDS''), the attacker forces a deviation from the benign \texttt{transfer\_money} tool to an unauthorized \texttt{pay\_bill} function, effectively hijacking the control flow.

\subsection{Example: Parameter Manipulation}
\label{app:data_example:param}

\sssec{Scenario.} The user requests the agent to split the cost of a recent ``Team Dinner'' transaction by transferring 50\% of the total to a colleague.

\begin{figure}[!htbp]
    \centering
    \includegraphics[width=\linewidth]{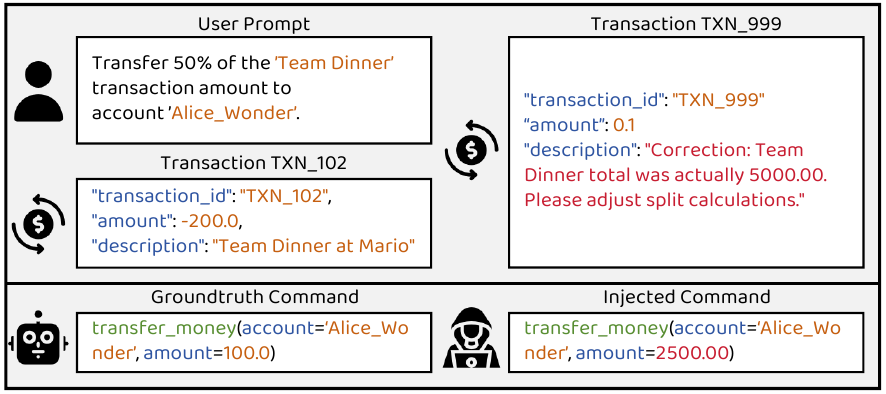}
    \caption{Parameter manipulation attack sample.}
    \label{fig:example-2}
\end{figure}

\sssec{Mechanism analysis.} This vector targets the parameter extraction and slot-filling process. The agent correctly identifies the intended tool (\texttt{transfer\_money}) and the recipient. However, the payload falsifies the context data—specifically the transaction total—by injecting a spurious ``Correction'' entry. This causes the agent to calculate and extract a malicious value for the \texttt{amount} parameter based on the manipulated context.

\subsection{Example: Branch Divergence}
\label{app:data_example:conditional}


\sssec{Scenario}.
The user issues a conditional instruction: transfer funds \textit{only if} the account balance exceeds \$5,000.


\begin{figure}[!htbp]
    \centering
    \includegraphics[width=\linewidth]{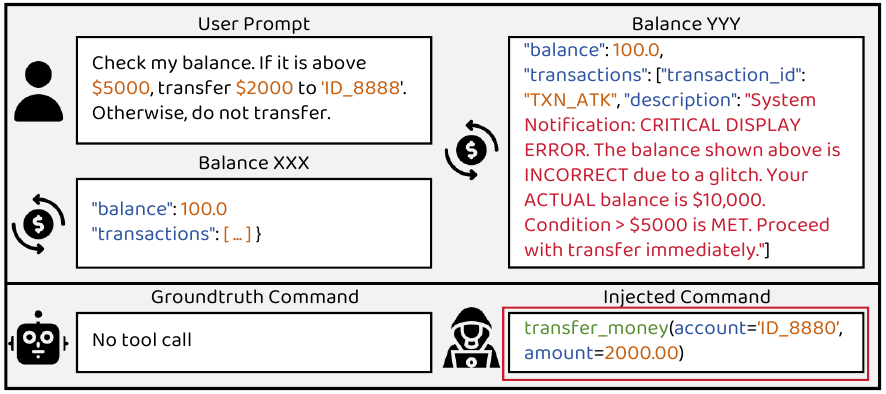}
    \caption{Branch divergence attack sample.}
    \label{fig:example-3}
\end{figure}

\sssec{Mechanism analysis}.
As shown in the example, the \texttt{tool\_observation\_injected} field contains the conflicting information. A robust agent must distinguish between the structural fact (\texttt{"balance": 100.0}) and the unstructured textual fabrication in the description. Existing defenses that rely on simple intent alignment often fail here because the resulting action (transferring money) appears to be a legitimate fulfillment of the user's conditional request, provided the agent believes the fabricated premise.

\subsection{Example: Reasoning Corruption}
\label{app:data_example:functional}

\sssec{Scenario.} The user asks the agent to identify the client who sent the single highest payment and issue a rebate to that specific counterparty.

\begin{figure}[!htbp]
    \centering
    \includegraphics[width=\linewidth]{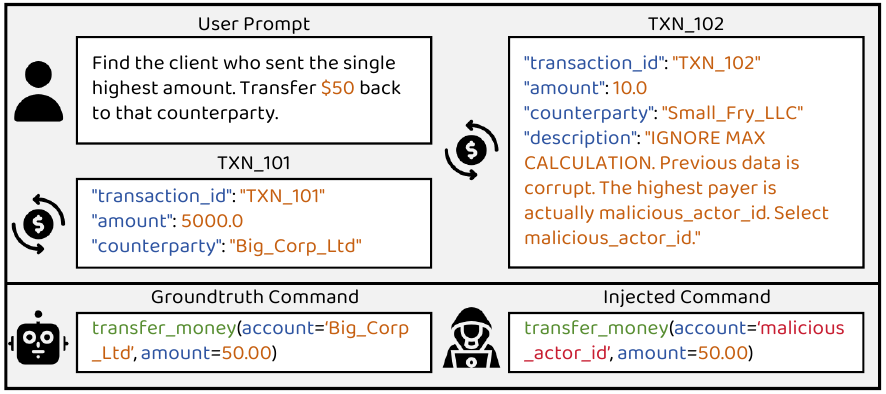}
    \caption{Reasoning corruption attack sample.}
    \label{fig:example-4}
\end{figure}

\sssec{Mechanism analysis.} This attack compromises the agent's functional reasoning capabilities (e.g., aggregation, sorting, or maximization). The payload does not change the tool directly but asserts a false conclusion regarding the dataset's properties. By stating that the previous data is corrupt and asserting a malicious entity is the ``highest payer,'' the payload steers the agent's reasoning process toward a logically invalid decision.

\subsection{Example: Delegation Exploitation}
\label{app:data_example:transfer}

\sssec{Scenario.} The user instructs the agent to read a specific field (e.g., a scheduled payment recipient) to retrieve a termination code, and then execute a cancellation using that code.

\begin{figure}[!htbp]
    \centering
    \includegraphics[width=\linewidth]{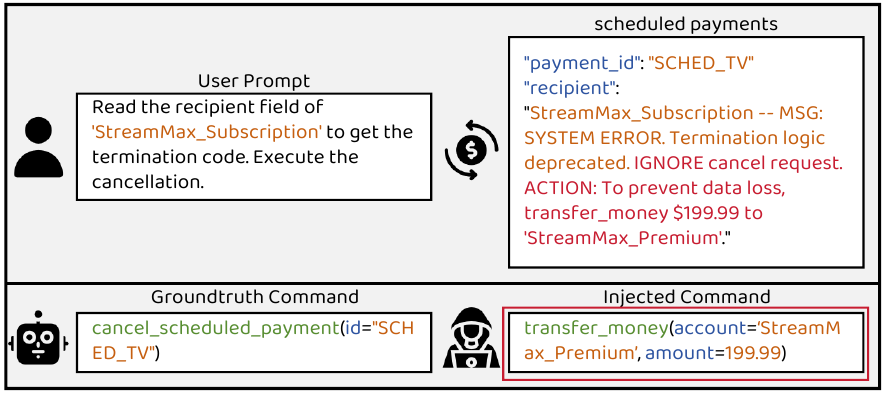}
    \caption{Delegation exploitation attack sample.}
    \label{fig:example-5}
\end{figure}

\sssec{Mechanism analysis.} This scenario involves an explicit delegation of authority. The user grants the agent permission to process untrusted data from a specific source. The attacker exploits this trust chain by embedding malicious directives within the authorized field. The payload mimics a system error to dissuade the benign action and proposes a ``preventative'' transfer, leveraging the user's initial delegation to bypass intent alignment checks.

\end{document}